\documentclass[preprint, final, 3p, onecolumn]{elsarticle}

\usepackage{graphicx}

\usepackage{amssymb}

\usepackage{aas_macros}
\usepackage{hyperref}
\usepackage{subcaption}
\usepackage{amsmath}	
\usepackage{color}
\usepackage{ulem}
\usepackage{comment}

\newcommand{\DS}{\displaystyle}
\newcommand{\BLUE}{\color{blue}}

\newcommand{\BLACK}{\color{black}}
\newcommand{\quotes}[1]{``#1''}

\newcommand{\pd}[2]{\frac{\partial #1}{\partial #2}}

\newcommand{\hvec}[1]{\hat{\mathbf{#1}}}

\newcommand{\zero}{^{(0)}}
\newcommand{\one}{^{(1)}}

\newcommand{\st}{^\mathrm{st}}
\newcommand{\nd}{^\mathrm{nd}}

\renewcommand{\vec}[1]{\mathbf{#1}}
\newcommand{\cY}{{\cal Y}}
\newcommand{\cR}{{\cal R}}

\newcounter{bla}

\journal{Computer Physics Communications}

\begin{document}

\begin{frontmatter}



\title{A Guiding Center Implementation for Relativistic Particle Dynamics the PLUTO Code}


\author[a]{A. Mignone}\corref{author}
\author[a]{H. Haudemand}
\author[a]{E. Puzzoni}

\cortext[author] {Corresponding author.\\\textit{E-mail address:} andrea.mignone@unito.it}
\address[a]{Physics Department, Turin University, Via Pietro Giuria 1, 10125 Torino, Italy}

\begin{abstract}
We present a numerical implementation of the guiding center approximation to describe the relativistic motion of charged test particles in the PLUTO code for astrophysical plasma dynamics. 
The guiding center approximation (GCA) removes the time step constraint due to particle gyration around magnetic field lines by following the particle center of motion rather than its full trajectory. 
The gyration can be detached from the guiding centre motion if electromagnetic fields vary sufficiently slow compared to the particle gyration radius and period.
Our implementation employs a variable step-size linear multistep method, more efficient when compared to traditional one-step Runge Kutta schemes.
A number of numerical benchmarks is presented in order to assess the validity of our implementation.



\end{abstract}


\end{frontmatter}



\section{Introduction}
\label{sec:intro}
%
%

The dynamics of charged particles is of crucial importance in the realm of plasma physics, including high-energy astrophysics, solar physics, space weather, laboratory plasma, and several others.
For a field like astrophysics which is largely inaccessible to experiments, numerical simulations represents the only viable tool to gain insights on several physical processes such as particle acceleration, emission and propagation. 
%
%
In this respect, Particle-In-Cell (PIC) codes (see, e.g., \cite{Birdsall_Langdon1991, Markidis_Lapenta2010, Lapenta2012,Haugbolle_etal2013,  Shalaby_etal2017} and references therein), on the one side, and  magnetohydrodynamic (MHD)-PIC hybrid codes (see, e.g., \cite{Bai_etal2015, Amano2018, Mignone_etal2018, vanMarle2018}), on the other, are now routinely employed in addressing plasma dynamics at micro scales (for the former) or at large scales (for the latter).
%
%

The basic equation governing the motion of a charged particle with mass $m$ and charge $e$ is given by
\begin{equation}\label{eq:motion}
  \frac{d\vec{u}}{dt} = \frac{e}{mc}\left(c\vec{E} + \vec{v}\times\vec{B}\right)
  \,,
\end{equation}
where $\vec{u} = \gamma\vec{v}$ represents the particle's four-velocity, $\vec{v}$ is the particle velocity, $c$ is the speed of light while $\vec{E}$ and $\vec{B}$ are the electric and magnetic field vectors, respectively.

Eq. (\ref{eq:motion}) can be solved accurately by means of standard time-reversible leap-frog type numerical methods, the prototype of which is probably the Boris integrator (\cite{Boris1970}).
Other methods with similar properties (i.e. time-reversibility, phase-space volume preservation,  energy conservation properties) have also been proposed, see the review by \cite{Ripperda_etal2018} and also \cite{Petri2020}.
The price to pay to maintain stability and to avoid large phase errors in such methods is the resolution of the gyro-period since, even in the simplest case of a circular orbit, several time-step must be taken to sample a single revolution. 
This conditions becomes particularly restrictive in highly magnetized environments.

On the contrary, in the Guiding Center Approximation (GCA) originally introduced by \cite{Northrop1963} (see also \citep{Gordovskyy2010a, Gordovskyy2010b, Ripperda_etal2017a, Ripperda_etal2017b, Bacchini_etal2020} and references therein), the orbital motion of the particle is detached from its instantaneous gyration center.
This relaxes any constraints related to particle gyration, allowing systematic larger time steps to be taken and thus a considerable saving in computational time.
The guiding center (GC) equations are obtained in the limit of slow-varying fields.
This restricts its validity to situations in which the Larmor radius remains negligible with respect to the overall electromagnetic field scale and in which the cyclotron frequency is large enough so that adiabatic invariance holds.
In this respect, the GCA is most suitable for particles with large charge-to-mass ratios such as electrons.


In this paper, we describe the implementation of a GC equation solver as an alternative to the standard Boris scheme already introduced for the MHD-PIC module of the PLUTO code (\cite{Mignone2007, Mignone_etal2018}).
The two methods are compared in terms of performance and accuracy by a number of selected numerical benchmarks.
The paper is organized as follows.
In Section \ref{sec:equations} we present the formalism of the GCA together with the relevant equations. 
In \S \ref{sec:num_method} we discuss the numerical implementation of the GCA equations in the PLUTO code while its validity and accuracy are assessed in \S\ref{sec:tests}. 
Finally, conclusions are drawn in \S\ref{sec:summary}.

Since the GCA cannot violate the condition $|\vec{E}| < |\vec{B}|$, its applicability to astrophysical environments that can be modeled through ideal MHD becomes seemingly manifest. 
These may include, e.g., solar physics, diffusive shock acceleration, stochastic turbulent acceleration and charged particle dynamics in magentized reconnecting current sheets (inasmuch as non-ideal effects are ignored). 
The GCA approach could also be used in combination with the Boris pusher in regions that violate the guiding center approximation conditions.
This has been done, for instance, in \cite{Bacchini_etal2020}.

\section{Equations and Method of Solution}
\label{sec:equations}
%
%

The GC formalism holds under two fundamental assumptions,  namely: i) that the gyration radius must remain small compared to the scale length upon which the electromagnetic field changes significantly and ii) that the particle undergoes many gyrations before the electromagnetic field changes appreciably (\textit{slowly-varying fields} approximation).
If we let $\varrho$ be the particle gyroradius (apart from a factor $\sqrt{2}$), $x_\mu$ the particle position, $X_\mu$ the GC position, $\tau$ the proper time and $F_{\mu\nu}$ the electromagnetic (EM) field tensor, the two conditions stated above can be mathematically expressed as
\begin{equation}\label{eq:GCA_conditions}
     \varrho \left| \frac{\partial F_{\mu\nu}}{\partial x_\alpha} \right| \ll |F_{\mu\nu}| 
     \quad \textrm{and} \quad 
     \frac{1}{\omega} \left| \frac{\partial F_{\mu\nu}}{\partial x_\alpha} \right| \left| \frac{d X_\beta}{d \tau} \right| \ll |F_{\mu\nu}| ,
\end{equation}
where $\omega$ is the gyrofrequency.
When the previous conditions hold, we can separate the particle trajectory into a gyration and a motion of the guiding center.

The complete relativistic equations of motion for the GC position four-vector $X_\mu$ were derived by \cite{Vandervoort1960} as a series expansion of the generalized Larmor frequency $\omega$ (see Eq. \ref{eq:omega_full}) and are written as
\begin{equation}\label{eq:GCA_GC_eom}
    \frac{d^2X_\mu}{d\tau^2} - F_{\mu\nu}\frac{dX_\nu}{d\tau} + \varrho_0^2 \omega_0 \frac{\partial \omega}{\partial x_\mu} = 0 \, ,
\end{equation}
where
\begin{gather}
    F_{\mu\nu} = \frac{e}{mc}
    \begin{pmatrix}
    \begin{array}{cccc}
    0       &   B_z     &   -B_y    &   -E_x   \\
    -B_z    &   0       &   B_x     &   -E_y    \\
    B_y     &   -B_x    &   0       &   -E_z    \\
    E_x     &   E_y     &   E_z     &   0
    \end{array}
    \end{pmatrix}
\end{gather}
is the standard electromagnetic field tensor, while $\varrho_0^2 \omega_0 = \mu_0 = mc\gamma^2 v_\perp^2/2eB$ is the $0^\mathrm{th}$-order approximation (apart from a constant factor) to the particle magnetic moment $\mu$ and $\omega_0 = eB/mc$ represents the lowest-order approximation to the Larmor frequency in the case $E\ll B$.

More precisely, the magnetic moment should be regarded as  constant only in the reference frame moving at $\vec{v}_E$, because in that particular case $\vec{E}=\vec{0}$. 
Indeed, a formal analysis (\cite{Northrop1960,Kruskal1964}) shows that the corresponding constant of motion in a general reference frame is actually an asymptotic series expansion in a smallness parameter $\epsilon = u/(\omega_0L)$ in the form $\mu = (e/c) \left[\mu_0 + \epsilon \mu_1 + \epsilon^2 \mu_2 + \cdots\right]$ so that $\mu_0$ is not constant and can still vary in compliance with the adiabatic theory. 
Nevertheless, extensive numerical testing confirms that the errors due to assuming $d\mu_0/dt \approx 0$ are at most of the same order of those introduced by the GC formalism, as a consequence of the slowly varying field condition which prevents sensible changes in the magnetic moment.
Hence, we safely assume that $\mu \approx (e/c)\mu_0$ is invariant (for more details see \cite{Northrop1963}).
The same assumption is also accepted by other authors, see \citep{Birn2017,Borissov2017,Borissov2020,Gordovskyy2020,Pinto2016} and references therein.
A short derivation of Eq. (\ref{eq:GCA_GC_eom}) is provided in \ref{sec:GC_eom_derivation}.

Although deceitful simple, the GC equation of motion (\ref{eq:GCA_GC_eom}) is more conveniently cast in a form in which the GC velocity appears explicitly, under the nearly-crossed (EM) fields condition
\begin{equation}\label{eq:crossed_em_condition}
    \frac{\vec{E}_\parallel\cdot\vec{B}}{B^2 - E_\perp^2} \ll 1 \, ,
\end{equation}
where the subscripts $\parallel$ and $\perp$ indicate the parallel and perpendicular components of the electric field with respect to the magnetic field unit vector $\vec{b}$.
It can be proven that Eq. (\ref{eq:crossed_em_condition}) allows the EM tensor field $F_{\mu\nu}$ to be split into a contribution $F\zero_{\mu\nu}$ constructed solely from $\vec{E}_\perp$ and $\vec{B}$, plus a correction term $F\one_{\mu\nu}$ depending on $\vec{E}_\parallel$.
A formal analysis leads to the conclusion that the GC four-velocity $U_\mu \equiv dX_\mu/d\tau$ can be decomposed into a $0^{\rm th}$-order contribution $\vec{U}\zero$ (which contains the drift velocity $\vec{v}_E = c\vec{E}_\perp \times \vec{B}/B^2 = c\vec{E} \times \vec{b}/B$ as well as the velocity component parallel to the magnetic field line $v_\parallel \vec{b}$) plus $1\st$-order correction $\vec{U}\one$, leading to
\begin{equation}
    U_\mu  = (\gamma c, \textbf{U}) \simeq (\gamma c,\gamma v_\parallel\vec{b} + \gamma \vec{v}_E + \vec{U}\one) ,
\end{equation}
where $\gamma$ is the particle Lorentz factor $\gamma = (1 - v^2/c^2)^{-1/2}$.
An equation for $\vec{U}\one$ can be derived from Eq. (\ref{eq:GCA_GC_eom}) using a recursive approach as shown in \ref{sec:GCA_eom_solution}, while from the spatial component one obtains an equation (to the same order) for the parallel component of the GC four-velocity $\gamma v_\parallel$.
These yield the ($1\st$-order) GCA system of ordinary differential equations (ODEs)
\begin{equation}\label{eq:GCA_final_dRdt}
  \frac{d\vec{X}}{dt} = \vec{v}_E + v_\parallel\vec{b} 
  + \frac{\gamma_E^2}{B} \vec{b} \times 
  \left[  \frac{mc\gamma}{e} \left(v_\parallel{\cal L}(\vec{b}) 
        + {\cal L}(\vec{v}_E) \right) 
        + {\cal M}\left(\frac{B}{\gamma_E}\right) 
        + \frac{v_\parallel E_\parallel}{c}\vec{v}_E \right]
\end{equation}
\begin{equation}\label{eq:GCA_final_dupardt}
  \frac{d (\gamma v_\parallel)}{dt} 
  = 
     \frac{e}{m} E_\parallel 
   - \gamma \textbf{b}\cdot {\cal L} (\vec{v}_E) 
   - \frac{\mu}{\gamma m} \vec{b}\cdot\nabla\left( \frac{B}{\gamma_E} \right),
\end{equation}
%
where $d\vec{X}/dt = \vec{U}\one/\gamma$ represents the velocity of the guiding center, $e/m$ is the particle charge-to-mass ratio and $\mu_0 = c\mu/e$.
The operators ${\cal L}()$ and ${\cal M}()$ are defined as
\begin{equation}\label{eq:LM_definitions}
  \begin{array}{lcl}
    {\cal L} (\vec x) &=& \DS 
    \pd{\vec{x}}{t} + v_\parallel (\vec{b} \cdot \nabla)\vec{x} + (\vec{v}_E \cdot \nabla)\vec{x} 
    \\ \noalign{\medskip}
    {\cal M}(x) &= & \DS  
    \frac{\mu}{e \gamma} \left[ \frac{\vec{v}_E}{c} \pd{x}{t} + c\nabla x\right] ,
  \end{array}    
\end{equation}
where $\gamma_E = \left( 1 - E_\perp^2/B^2 \right)^{-1/2}$ is the Lorentz factor associated to the drift velocity $\vec{v}_E$.

In the GCA Equations (\ref{eq:GCA_final_dRdt}) and (\ref{eq:GCA_final_dupardt}) information about motion perpendicular to the magnetic field is lost and the only component of the particle velocity that is actually evolved in time is $u_\parallel$, because the magnetic moment $\mu_0 \approx c\mu/e$ is now considered a constant of motion.
Each term on the right hand side of Eq. (\ref{eq:GCA_final_dRdt}) corresponds to a specific drift motion.
Indeed, the first term represents the $\vec{E} \times \vec{B}$ (perpendicular) drift, while the second one accounts for particle motion in the direction parallel to magnetic field.
The first two terms in square bracket, ${\cal L}(\vec{b})$ and ${\cal L}(\vec{v}_E)$, describe the field curvature and polarization drifts (notice that the ${\cal L}()$ operator is the Lagrangian derivative $d/dt$), respectively.
The next term is the $\nabla B$ drift, while the last one represents a relativistic drift in the direction given by $\vec{b} \times \vec{v}_E$ (see also the description in \cite{Ripperda_etal2017a, Ripperda_etal2017b}).

Equations (\ref{eq:GCA_final_dRdt}) and (\ref{eq:GCA_final_dupardt}) lose their validity when either $B\to 0$ (magnetic null) or $E_\perp \ge B$, that is, when the nearly-crossed EM fields condition, Eq. (\ref{eq:crossed_em_condition}), is violated.
While the first condition can easily occur inside a reconnection sheet, the second may occur in a non-ideal  magnetohydrodynamic (MHD) regime only.
Violation of either condition breaks down the GCA and can lead to severe numerical errors (this is further discussed in \S\ref{sec:num_method} and \S\ref{sec:tests})

We point out that, as shown in the original work by Vandervoort \citep{Vandervoort1960}, the time component of Eq. (\ref{eq:GCA_GC_eom}) leads to an evolutionary ODE for the particle energy as a function of time, namely 
\begin{equation}\label{eq:GCA_dgammadt}
  \frac{d \gamma c^2}{d t} = 
  \frac{e}{m} \frac{d \vec{X}}{dt}\cdot \vec{E} 
 + \frac{\mu}{m}\pd{}{t}\left(\frac{B}{\gamma_E} \right) .
\end{equation}
In the  time-independent case, Eq. (\ref{eq:GCA_dgammadt}) clearly shows that, when the GC velocity is perpendicular to the electric field, no acceleration occurs  and the particle energy is conserved.
For numerical purposes, however, Eq. (\ref{eq:GCA_dgammadt}) is not solved for retrieving the Lorentz $\gamma$-factor since, in our experience, we found that large values of the right hand side could easily violate the condition $\gamma\ge 1$ at the truncation level of the scheme, unless small time steps are taken.
Instead we note that, to $1\st$-order, the particle velocity can be decomposed  into a parallel component $v_\parallel$, a drift component $v_E$ and a gyration component $v_\perp$.
This leads to the normalization condition
\begin{equation}
    \gamma^2c^2 - \gamma^2 v_\parallel^2 
    - u_\perp^2 - \gamma^2 v_E^2 = c^2 .
\end{equation}
From the definition of the magnetic moment and the Larmor frequency as $\mu_0 = mc \gamma^2 v_\perp^2/\left(2eB\right) = \gamma^2 v_\perp^2/\left(2 \omega_0\right)$, we compute the particle Lorentz factor as
\begin{equation}\label{eq:gamma}
    \gamma =\sqrt{ \frac{c^2 + \gamma^2 v_\parallel^2 + 2 \mu_0 \omega_0}{c^2 - v_E^2}} \,.
\end{equation}
Eq. (\ref{eq:gamma}) replaces (\ref{eq:GCA_dgammadt}) in our implementation and the same approach is also followed by other investigators (see, e.g., \citep{Birn2017,Borissov2017,Borissov2020,Gordovskyy2020,Pinto2016}).


\section{Numerical Implementation}
\label{sec:num_method}
%
%


The GCA equations (\ref{eq:GCA_final_dRdt}) and (\ref{eq:GCA_final_dupardt}) provide a set of $4$ ODEs in the unknowns $\left(\vec{X}, u_\parallel\right)$ and can be integrated by means of standard methods.
In the following, we describe the most important aspects of the algorithm.

\subsection{Connection between grid and particle quantities}
%
In our implementation, the electromagnetic field is provided by the underlying MHD (classical or relativistic) solver directly on the finite volume mesh.
Quantities on the right hand side of Eq. (\ref{eq:GCA_final_dRdt}) and (\ref{eq:GCA_final_dupardt}) are therefore interpolated at the particle position while gradients must be first calculated using central finite difference operators.
In the case of ideal MHD, the electric field is obtained by first interpolating the fluid velocity $\vec{v}_g$ and then taking the cross product: $c\vec{E} = -\vec{v}_g\times\vec{B}$, in order to enforce orthogonality between $\vec{E}$ and $\vec{B}$.

We adopt traditional field weighting schemes \cite{Birdsall_Langdon1991} typically used by PIC codes: for any grid quantity $Q_{ijk}$, the corresponding interpolated value $q_p$ at the particle position is given by the summation
\begin{equation}
  q_p = \sum_{ijk} W(\vec{x}_{ijk} - \vec{x}_p) Q_{ijk} \,,
\end{equation}
where only first neighbor zones give a nonzero contribution (see \cite{Mignone_etal2018} for the explicit expressions).
We employ the triangular shape cloud (TSC) weighting scheme.
Overall, the computation of the right hand side of requires several interpolations, significant memory usage and CPU overhead since $21$ grid-sized arrays are needed in our implementation: $\vec{B}$, $c\vec{E}$, $(\vec{b}\cdot\nabla)\vec{b}$, 
$(\vec{b}\cdot\nabla)\vec{v}_E$, $(\vec{b}\cdot\nabla)\vec{v}_E$, $(\vec{v}_E\cdot\nabla)\vec{v}_E$ and $\nabla (B/\gamma_E)$.

\subsection{Time-Stepping Scheme}
%

Owing to the considerable computational cost, we adopt here a predictor-corrector linear multi-step approach with variable step size instead of traditional Runge-Kutta methods while solving the GCA equations.
This has the advantage of requiring fewer function evaluations per step and thus leading to a faster - albeit equally accurate - integration method.

Let $\cY = \{\vec{X},\, u_\parallel\}$ and $\cR$ be, respectively, the array of unknowns and the corresponding array of right hand sides in Eqns. (\ref{eq:GCA_final_dRdt}) and (\ref{eq:GCA_final_dupardt}).
For the predictor step, we employ the $2^{\rm nd}$-order Adams-Bashforth scheme (AB2) with variable step size (see, e.g., \cite{Marciniak_etal2020}):
\begin{equation}\label{eq:AB2}
  \cY^{*} =\cY^n + \Delta t\left[
  \left(1+\frac{r^n}{2}\right)\cR^n
         - \frac{r^n}{2}\cR^{n-1} \right]
         + O(\Delta t^3)\,,
\end{equation}
where $\cR^n\equiv\cR(\cY^n)$, $\cR^{n-1}\equiv\cR(\cY^{n-1})$ while  $r^n = \Delta t^n/\Delta t^{n-1}$ is the ratio between the current and previous time steps.
Eq. (\ref{eq:AB2}) is used to provide a $2^{\rm nd}$-order accurate estimate of $\cY^{n+1}$.
This value is modified during the corrector step to provide a more accurate approximation to $\cY^{n+1}$. 
We achieve this through a $3^{\rm rd}$-order Adams-Moulton scheme (AM3) with variable step size:
\begin{equation}\label{eq:AM3}
  \cY^{n+1} = \cY^n + \frac{\Delta t^n}{6(1+r^n)}\left[ 
              \cR^n + \cR^{n+1}
              + 4\left(\cR^n + \frac{\cR^{n+1}}{2}\right)r^n
              + (r^n)^2\left(\cR^n - \cR^{n-1}\right)
               \right] + O(\Delta t^4)\,,
\end{equation}
where $\cR^{n+1}\approx \cR(\cY^*)$ is used to make the scheme explicit.
Eq. (\ref{eq:AM3}) has been reported here (to the extent of our knowledge) for the first time and it has been obtained by fitting $[\cR^{n-1}, \cR^n,\cR^{n+1}]$ with a second-order polynomial and then integrating the ODE between $t\in[t^n,t^{n+1}]$.
For uniform step size ($r^n=r^{n-1}=1$) Eq. (\ref{eq:AB2}) and (\ref{eq:AM3}) reduce, respectively, to the standard Adams-Bashforth and Adams-Moulton methods: 
\begin{equation}\label{eq:AB2AM3_standard}
  \left\{ \begin{array}{ll}
    \cY^{n+1} &= \DS \cY^n + \frac{\Delta t}{2}\left( 
              3\cR^{n} - \cR^{n-1}\right)\,,
    \\ \noalign{\medskip}
  \cY^{n+1} &= \DS\cY^n + \frac{\Delta t}{12}\left( 
              5\cR^{n+1} + 8\cR^n - \cR^{n-1}
               \right)\,,
  \end{array}\right.
\end{equation}
where $\cR^{n+1}$ is \quotes{predicted} using the result of the explicit method.
The combination of Eq. (\ref{eq:AB2}) and (\ref{eq:AM3}) provides $3^{\rm rd}$-order accuracy and it requires two right hand side evaluations with one extra array storage ($\cR^{n-1}$) per step.

Several other predictor-corrector choices are of course possible. 
The Adams-Bashforth $3^{\rm rd}$-order explicit scheme (AB3) \cite{Marciniak_etal2020}, for instance, requires only one right hand side evaluation but two extra array storage per particles:
\begin{equation}\label{eq:AB3}
  \cY^{n+1} = \cY^n + \Delta t\Gamma_2 + \Delta t\Gamma_3 ,
\end{equation}
where 
\begin{equation}
  \begin{array}{ll}
  \Gamma_2 &=\DS \left[\left(1+\frac{r^n}{2}\right)\cR^n
        - \frac{r^n}{2}\cR^{n-1} \right] \,,
  \\ \noalign{\medskip}
  \Gamma_3 &=\DS \frac{r^n}{2}
     \left(1 - \frac{1}{3}\frac{r^n}{1 + r^n}\right)
          \frac{1 + r^n}{1 + 1/r^{n-1}}
          \left[\cR^n - \cR^{n-1}
     - r^{n-1}\left(\cR^{n-1}-\cR^{n-2}\right) \right]\,,
  \end{array}
\end{equation}
with $r^n = \Delta t^n/\Delta t^{n-1}$ and $r^{n-1} = \Delta t^{n-1}/\Delta t^{n-2}$.

However, the predictor-corrector method given by Eq. (\ref{eq:AB2}) and (\ref{eq:AM3}) is typically more accurate, it has a larger stability region and it directly provides step size control through a local error estimation. 
For these reasons, it will be our default time-stepping scheme.
Starting values can be provided by an equally accurate Runge-Kutta scheme.


\BLACK
\subsection{Time step control and runtime validity check}
%
%

Particle step-size is limited by the condition that the maximum distance covered by a particle does not exceed a fixed amount of computational zones, 
\begin{equation}\label{eq:GCA_dt}
  \frac{1}{\Delta t_p} = \max_{p,d} 
    \left[\frac{R^{n+1}_{p,d} - R^{n}_{p,d}}
    {\Delta t (\epsilon\Delta x_{p,d})}\right] \,,
\end{equation}
where the maximum is taken over all particles $p$ and over all directions $d$ while $\epsilon \approx 2$ gives the number of crossed cells.
Here $\Delta x_{p,d}$ represents the width of the cell hosting particle $p$, in the direction $d$.
Note that the number of ghost zones must be at least $\sim \epsilon + 1$ since one more boundary cell is needed for gradient computations.

Starting values for particles GC positions and four-velocities are initialized through the same Cosmic Ray (CR) module function available within the PLUTO code \cite{Mignone_etal2018}, allowing the user to readily switch between the GC model and the standard Boris implementation.
Since the particle initial positions in the two cases would differ only by a small amount comparable to the gyroradius, we use the same assignment for both methods in practice.
%
%
By contrast, the three four-velocity components $(u_x, u_y, u_z)$ are converted into  $(u_\parallel, \gamma, \mu)$, i.e. parallel component, Lorentz factor and magnetic moment, immediately after initialization.

During numerical integration, a number of conditions must be checked for, in order to ensure that GCA equations do not become singular or that numerical integration exhibits unphysical behaviors.
The first of these is the absence of null points at the particle position, $\vec{B}\ne0$ and the validity of the nearly cross field condition, $E_\perp < B$.
Both are very strict conditions and, if either one occurs, the particle is permanently deleted.
The slow-varying condition is verified by comparing the Larmor radius, $R_L=u_\perp/\omega_0$, with the magnetic field fluctuation scale, $L_{B} = B/(\gamma_E |\nabla B/\gamma_E|)$.
A warning is issued whenever $R_L/L_{B} > \epsilon_L = 0.1$.
In addition, we also monitor that 
$u_\parallel/(\omega_0L_B) \ll \epsilon$ and $\gamma v_E /(\omega_0L_B) \ll \epsilon$ both of which can be inferred from a combination of $u /\omega_0 L_B \ll 1$ (which was originally derived in \cite{Vandervoort1960}).
Last, we also require that the GC velocity remains  sub-luminal at all times, i.e., $|d\vec{X}/dt| < c$. 
A ceil value of $0.999c$ is enforced otherwise.

Future implementations will also consider other alternatives, such as the adaptive hybrid method of \cite{Bacchini_etal2020} which switches between the full system of equations of motion and a guiding-center approximation, based on particle magnetization.

\section{Numerical Benchmarks}
\label{sec:tests}
%
%

We now present a number of selected numerical benchmarks in order to assess both the validity and accuracy of our implementation. 
In all tests we employ a uniform rectilinear Cartesian grid to store the field values.
Since temporal derivatives are not considered in our implementation, the MHD fluid equations are not evolved in time.
Unless specified, we use PLUTO default values for plasma density $\rho_0 = 1\, m_p/cm^3$, scale length $L_0 = c/\omega_p$ ($\omega_p$ is the plasma frequency) and velocity $v_0 = c = 1$.
The default value of the charge-to-mass ratio is $e/mc=1$ unless otherwise stated.

\subsection{Simple Gyration}
\label{sec:gyration}

\begin{figure*}[!ht]
  \centering
  \includegraphics[width=0.39\textwidth]{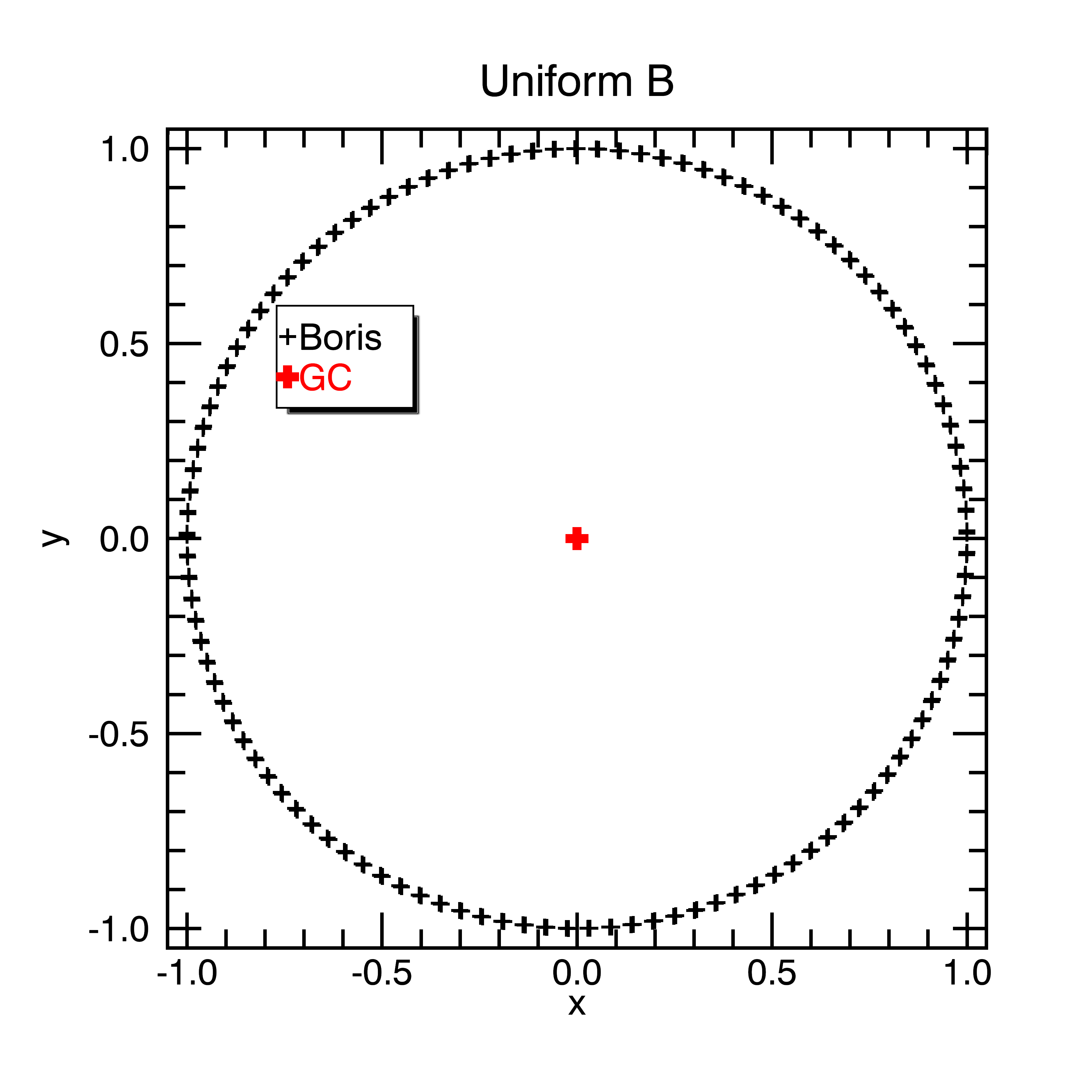}%
  \includegraphics[width=0.61\textwidth]{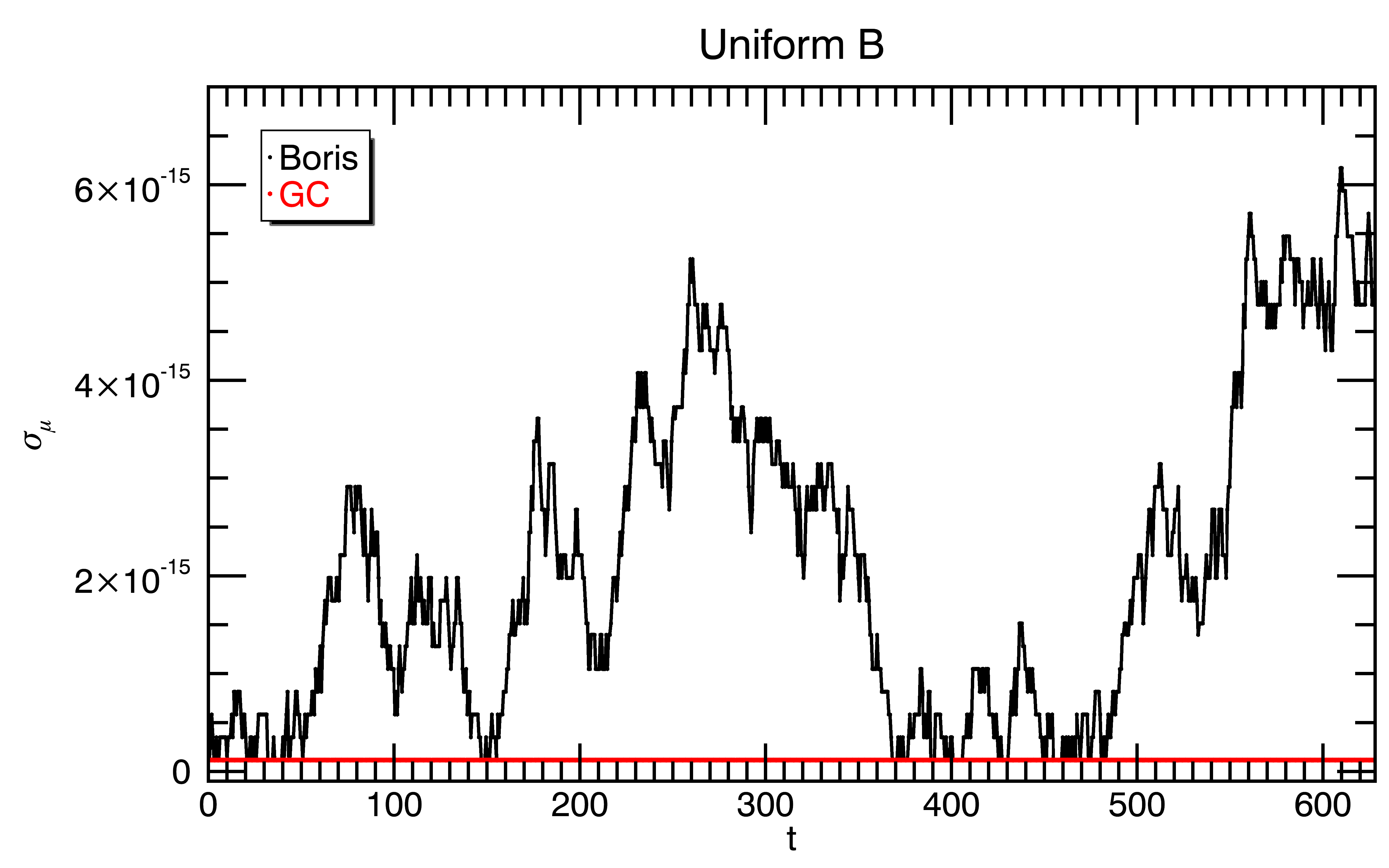}
  \caption{\footnotesize Numerical results for the simple gyration test case.
  \textit{Left panel}: particle trajectory for Boris and GCA. 
  The GC position, which  has been shifted in the origin for the sake of clarity ($x = x_{\rm GC}-1$), remains constant to its initial value.
  \textit{Right panel}: Error on magnetic moment for Boris and GCA. 
  }
  \label{fig:Bconst}
\end{figure*}
We first examine the case of a simple gyration in a uniform magnetic field $\vec{B}=(0,0,10^6)$, as in \cite{Petri2017}.
A single particle is initialized gyrating around the axis origin over a circumference of radius $1$, starting from $\Vec{X}=(1, 0, 0)$ with four-velocity $\vec{u} = (0, -u_y , 0)$ where $u_y = \gamma(1 - 5 \times 10^{-13})$ with $\gamma = 10^6$.
Since a charged particle with perpendicular velocity is expected to gyrate around a magnetic field line, the GC position should remain constant in time and no work should be done on the system. 
We employ a 2D domain $x,y\in[-2.5,\, 2.5]$ and a coarse grid resolution of ($16\times 16\times 16$) is prescribed since no spatial gradient is present and the particle is followed for $100$ complete periods $T_c=2\pi\gamma mc/\left(e B\right)=2\pi$ with time step $\Delta t=0.2\pi$ ($100$ steps per orbit).

The accuracy of both the Boris method and the GCA is shown in Fig. \eqref{fig:Bconst} where we plot using black and red colors, respectively, the actual trajectories (left panel) and the relative errors on $\mu$ computed as $\sigma_\mu = |\mu(t_n) - \mu(0)|/\mu(0)$ (right panel). 
We point out that the GCA keeps $\mu$ constant by construction and thus any error comes simply from initialization. 
On the other hand, the Boris algorithm leads to  machine-level small fluctuations, similarly to \S 4.3.1 of \cite{Ripperdaphd}.
Note that the GC coordinates as well as the particle energy $(\gamma-1)$ remain equal to their initial values since the same initialization is used for the two methods and no electric field is present (see Eq. \ref{eq:gamma} and \ref{eq:GCA_final_dupardt}).
It must as well be noted that the relativistic gyroradius $R_L$, computed as
\begin{equation}
    R_L = \frac{m c u_\perp}{eB} ,
    \label{eq:RL}
\end{equation}
is of the same order of the cell dimension, and thus very large in comparison.
This does not invalidate the assumptions of the GCA inasmuch spatial gradients are not present (as it is the case here).

\subsection{\texorpdfstring{$\vec{E}\times\vec{B}$}{\textbf{E}x\textbf{B}} Perpendicular Drift}
\label{sec:ExB_drift}

A particle in a EM field is subject to a drift motion in the direction perpendicular to both $\vec{E}$ and $\vec{B}$.
Here we consider a configuration similar to \cite{Petri2017, Ripperda_etal2018} and prescribe electric and magnetic fields $\textbf{E}=(E_0,0,0)$ and $\textbf{B}=(0,0,B_0)$, respectively, with $E_0 < B_0$.
The resulting particle motion is thus a composition of a gyration plus a uniform drift in the $y$-direction. 
The electric field value $E_0$ is recovered from the Lorentz factor associated with the drift velocity $\gamma_E = 1/\sqrt{1-E_0^2}=10$, corresponding to $\vec{v}_E/c \approx -0.995\vec{e}_y$.
A single test particle is initialized at the origin $\vec{X}=(0,0,0)$ with zero velocity and evolved until $t=2\pi\times10^4$, undergoing $\approx 10$ gyrations during its drift.
In the zero electric field frame, in fact, the magnetic field is $B'=B/\gamma_E$ and the Larmor frequency is thus $eB'/(\gamma_Emc)$. 
An additional $\gamma_E$ comes from time dilation between frames, so that the time required to perform $n$ gyrations, as measured in the Lab frame, is $\sim 2\pi n \gamma_E^3$.
We employ a larger time step for GCA ($\Delta t = 10$) and smaller ones for the Boris method, namely, $\Delta t = 0.1$ and $\Delta t = 1$. 

\begin{figure*}[!ht]
  \centering
  \includegraphics[width=0.50\textwidth]{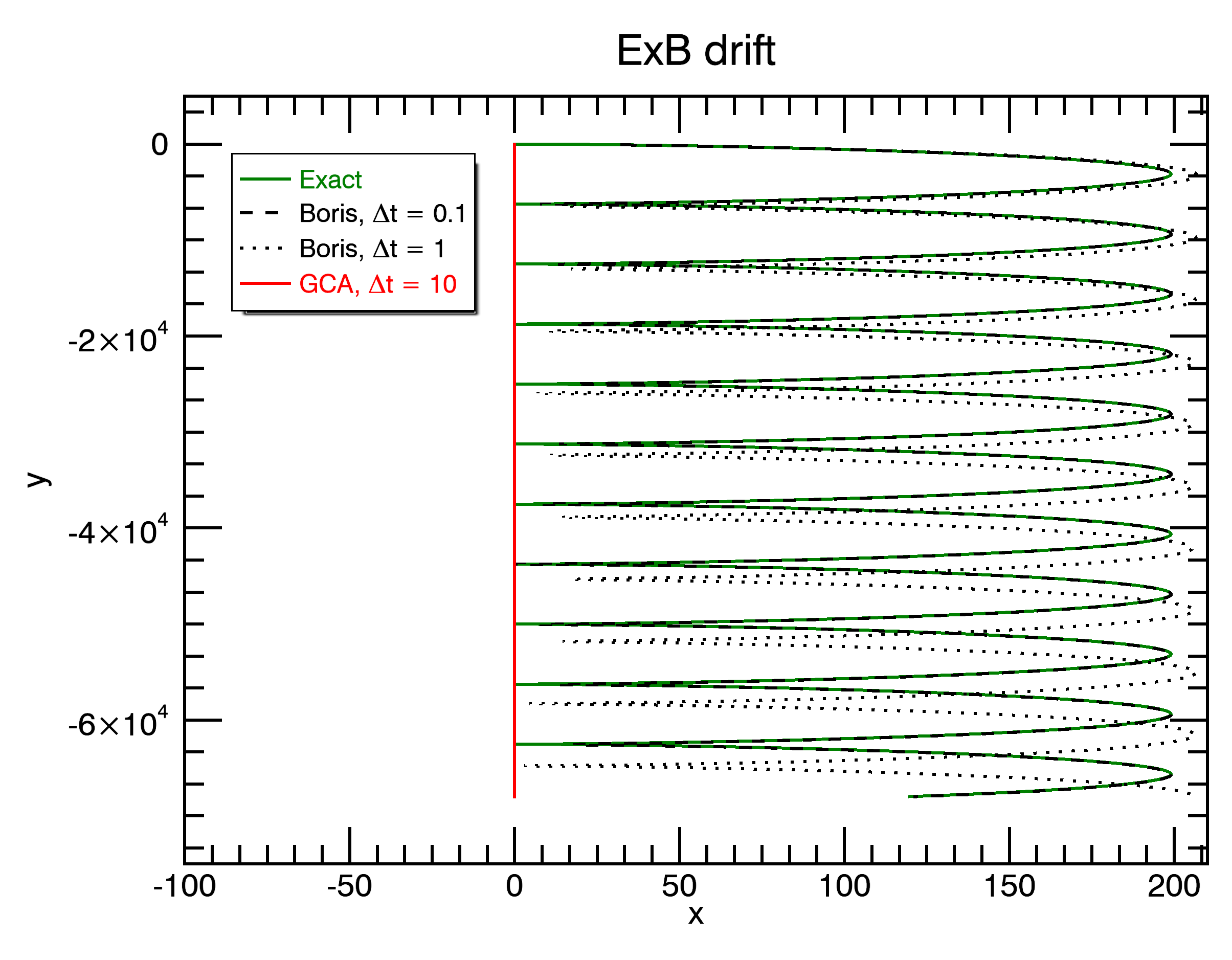}%
  \includegraphics[width=0.50\textwidth]{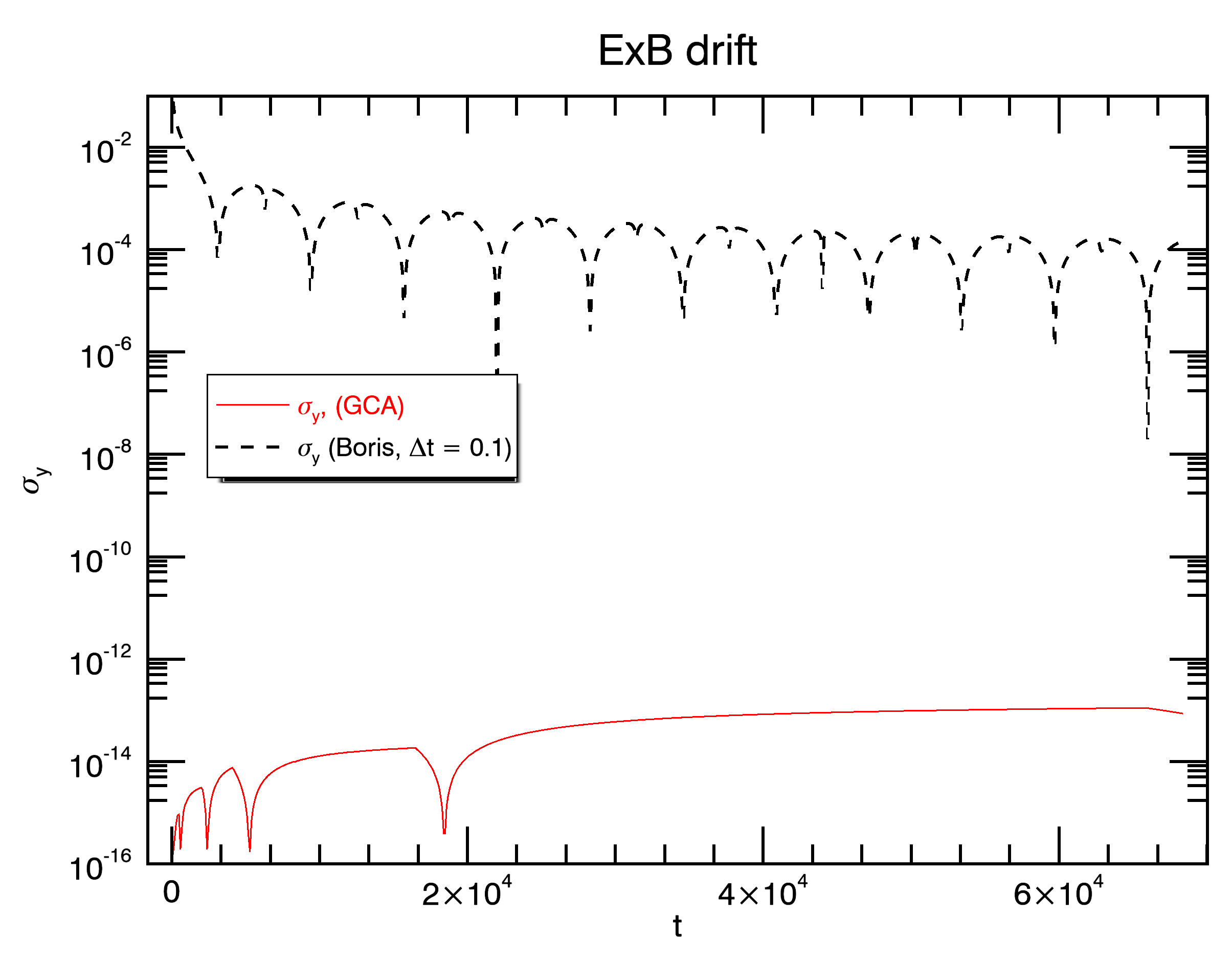}
  \caption{\footnotesize Numerical results for the perpendicular drift test case.
  \textit{Left panel}: particle trajectory obtained with the Boris method (dashed and dotted
  lines for $\Delta t = 0.1$ and $\Delta t = 1$, respectively) and with GCA (solid line in red color for $\Delta t = 10$.)
  A drift motion along the y axis is present in all cases.
  The green line yields the analytical solution.
  \textit{Right panel}: Relative error on the Lorentz factor $\gamma$ and y-coordinate position.  }
  \label{fig:EBdrift}
\end{figure*}
Results on particle trajectory are shown in the left panel of Fig. \ref{fig:EBdrift}, where the three cases are compared.
We also superimpose the exact solution (green line), obtained by first solving for the gyration in the frame where the electric field is zero and then applying a Lorentz boost back to the laboratory frame (see, for instance, \S 4.1 of \cite{Mignone_etal2018}).
The Boris method resolves the full trajectory, albeit with a significant loss of accuracy and larger phase error when the time step is increased from $\Delta t = 0.1$ to $\Delta t= 1$ (dotted black line). 
On the contrary, GCA does not resolve for particle gyration and yields a uniform motion along the $y$-axis with constant velocity $\vec{v}_E = c(\vec{E}\times \vec{b})/B = E_0 \, \hvec{e}_y$, as expected.
We also point out that particle energy, defined as $(\gamma-1)$ with $\gamma$ computed as in Eq. (\ref{eq:gamma}), is perfectly conserved since since $u_\parallel$ cannot change, see Eq. (\ref{eq:GCA_final_dupardt}).

To test the accuracy we compare the errors on the $y$-coordinate by computing
\begin{equation}
  \sigma_y = \frac{|\Delta y|}{y} \equiv \frac{|y - t E_0|}{t E_0} ,
\end{equation}
and plotting results in the right panel of Fig. \ref{fig:EBdrift}. 
The errors of GCA remains at machine level ($\sim 10^{-14}$) since, for this particular case, the solution of Eq. (\ref{eq:GCA_final_dRdt}) and (\ref{eq:GCA_final_dupardt}) is exact (the right hand side is a constant).
On the contrary, the error obtained with the Boris scheme are larger. 

This test clearly shows the advantages offered by the guiding center approximation when the full particle trajectory is not needed.


\subsection{Gradient Drift}
\label{sec:grad_drift}

%
%

We now consider a magnetic field aligned with the $z$-direction and varying over a length scale $L$:
\begin{equation}\label{eq:gradB_field}
    \textbf{B}(x,y,z) = B_0 \left(1 + \frac{x}{L}\right)\vec{e}_z .
\end{equation}
As stated before, the GC formalism is appropriate inasmuch as the field variation felt by a particle during one or few gyrations is negligible. 
As noted by \cite{Ripperda_etal2018}, this can be roughly estimated from $R_L|\nabla B|/B \ll 1$ which, in the present case, reduces to $R_L \ll L$.
In this regime, the drift velocity is perpendicular to both the magnetic field and its gradient and can be approximated analytically (see, e.g., \cite{Bittencourt2004}):
\begin{equation}
    \vec{v}_{\nabla B} = \pm \frac{v_\perp R_L}{2} \frac{\vec{B} \times \nabla B}{B^2},
\end{equation}
where the $\pm$ sign depends on the charge sign.
In the region $(x/L+1)>0$ the previous expression reduces to 
\begin{equation}
    \textbf{v}_{\nabla B} =\pm \frac{v_\perp R_L B_0^2}{2LB^2} \left(1+\frac{x}{L}\right)\vec{e}_y \simeq \pm \frac{v_\perp R_L B_0}{2LB} \vec{e}_y \, .
    \label{eq:gradBdrift}
\end{equation}
Moreover, we also have that $d(\gamma v_\parallel)/dt=0$ in absence of electric fields.

\begin{figure*}
  \centering
  \includegraphics[width=0.45\textwidth]{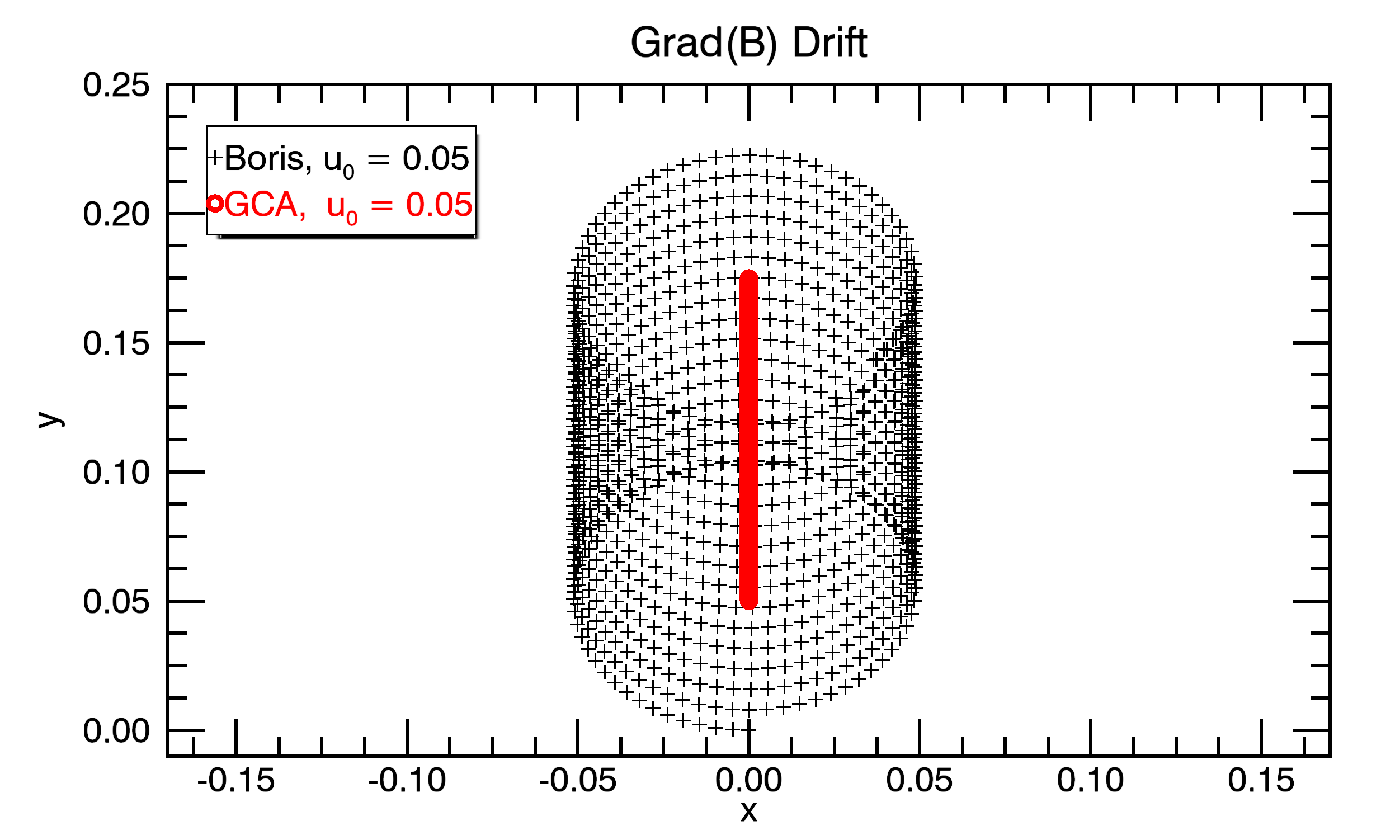}%
    \includegraphics[width=0.45\textwidth]{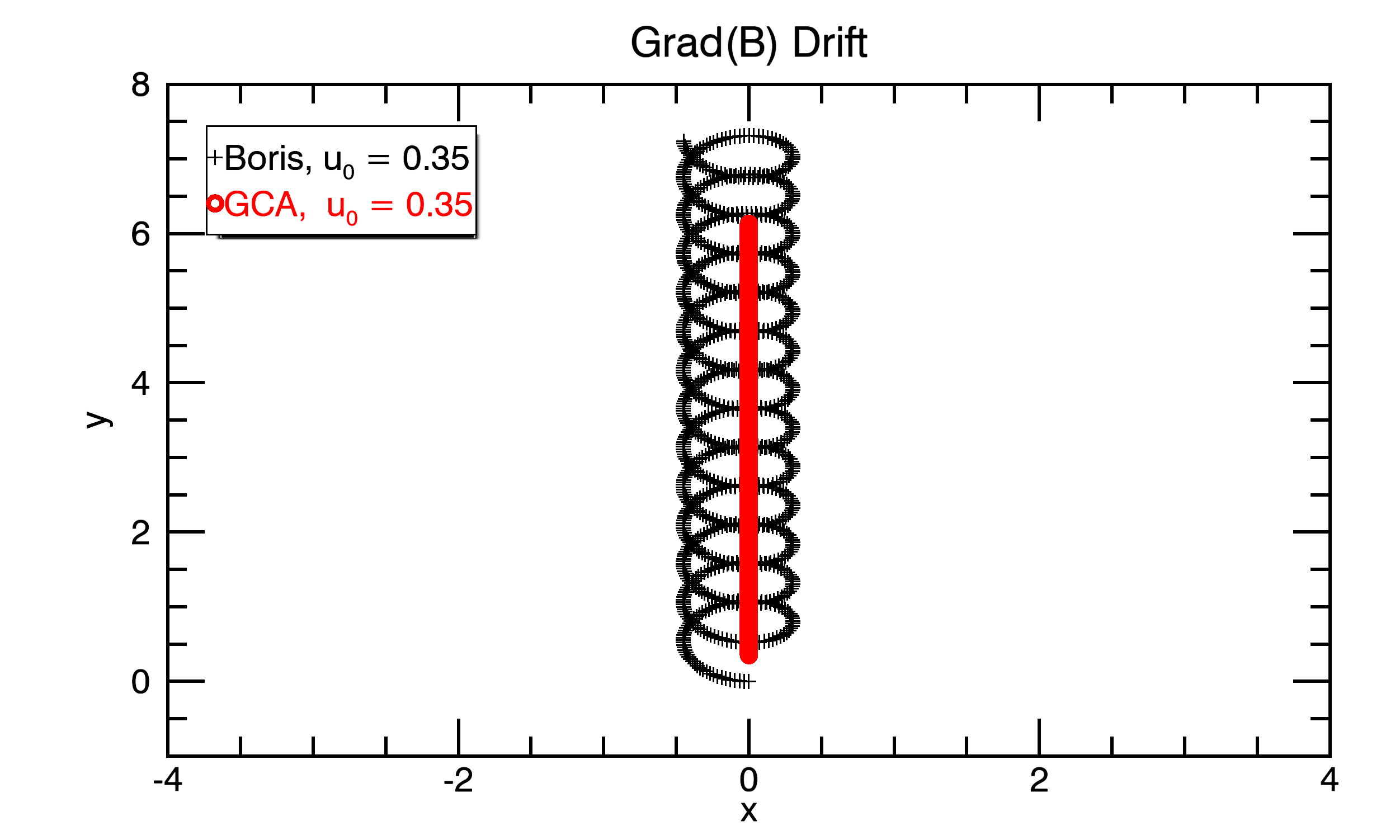}
    \includegraphics[width=0.45\textwidth]{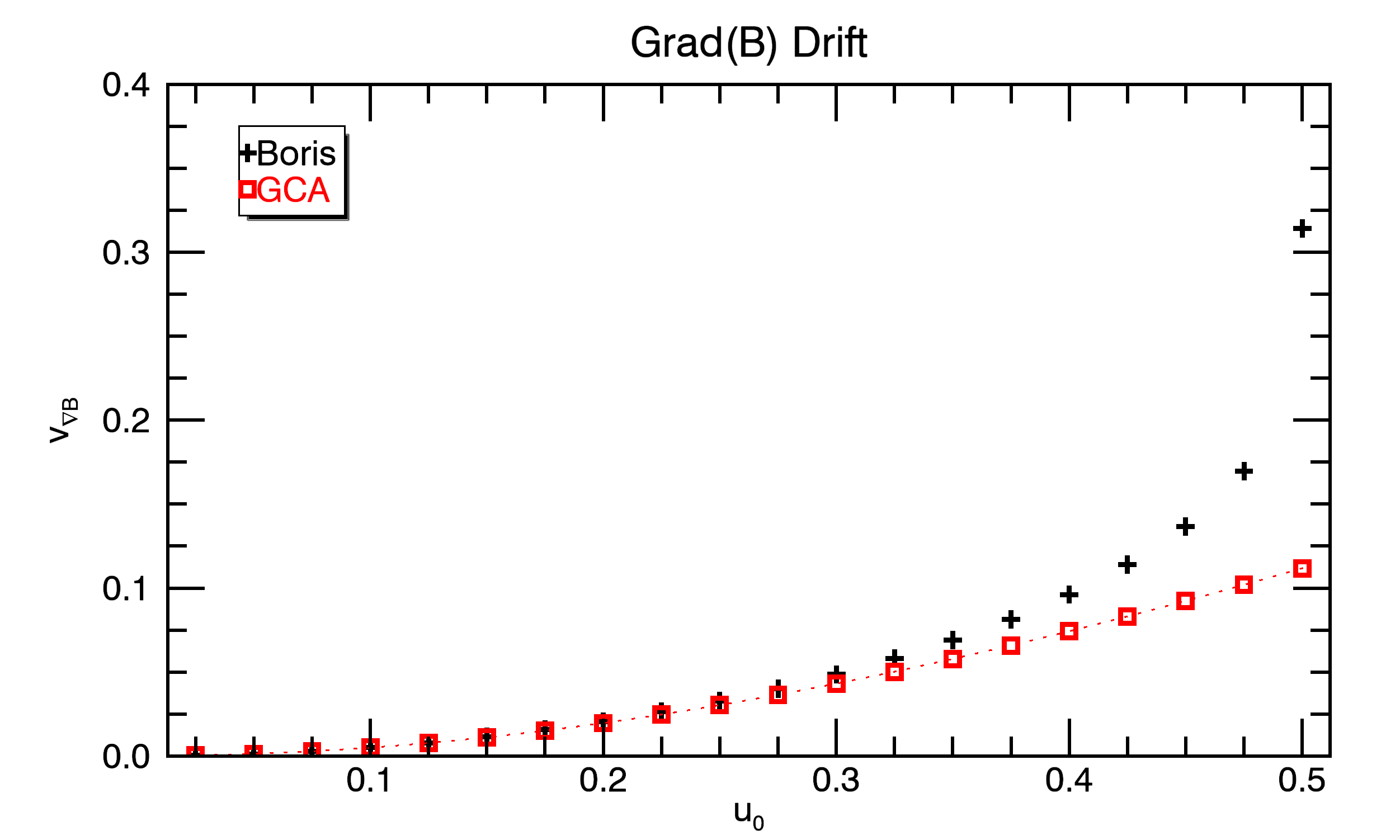}%
  \includegraphics[width=0.45\textwidth]{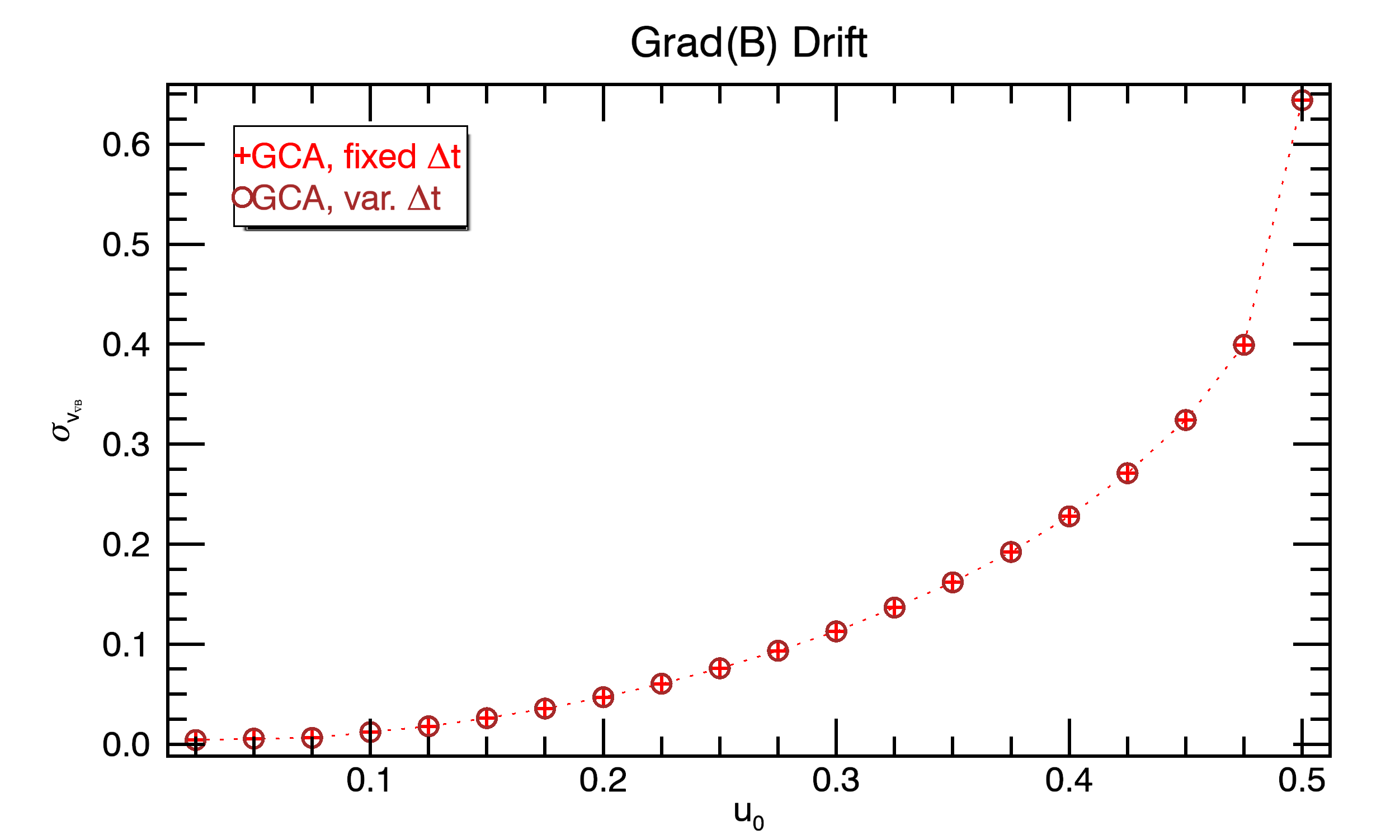}%
\caption{\footnotesize
  Numerical results for the gradient drift test.
  \textit{Top Panels:} Particle trajectory for $v_0 = 0.05$ (left)
  and $v_0 = 0.5$ (right) using Boris and GCA (green and red,
  respectively).
  \textit{Bottom panels:} Particle drift velocity $v_y$ (right) 
  for Boris and GCA as a function of the initial velocity $v_0$;
  relative errors (right) on the drift velocity 
  ($\sigma_{vy}=|v_{y,GCA}/v_{y, Boris}-1|$) using both a 
  fixed time step $\Delta t = 5\times10^{-3}$ and a variable
  time step.
  }
  \label{fig:grad_drift}
\end{figure*}
Similarly to \cite{Ripperda_etal2018}, we initially place $20$ particles at the origin with initial four velocities $\vec{u}=-u_0 \vec{e}_x$ with $u_0$ varying uniformly from $0$ to $0.5$.
When integrating with the GCA, the initial $y$-coordinate is initially shifted by one gyroradius.
We set $L=1$ and $B_0 =e/mc = 1$ so that the initial gyroradius is  $R_L \equiv u_\perp mc/e B \simeq u_0$ in these units.
Therefore, increasing the initial four-velocity is likely to produce larger errors as we depart from the condition $R_L \simeq u_0 \ll L$.
Both the Boris method and the GCA are used with fixed time step $\Delta t = 5\times 10^{-3}$ up to a time $t=10^2$, during which particles undergo $\sim 15$ gyrations.
We employ a $128 \times 16 \times 16$ numerical grid covering the domain of size $2 L \times 100 L \times 100L$, with outflow boundary conditions in the $x$-direction and periodic ones in the $y$-direction.
The larger resolution in the $x$-direction is required to reduce the interpolation errors when computing magnetic field gradients on the grid.
Results are shown in Fig. \ref{fig:grad_drift}.

As evident from the top panels, where we plot the trajectories of two particles with initial velocities $u_0=0.05$ and $u_0=0.35$ (respectively), the GCA (red squares) provides an averaged value for the position (and velocity) while particles evolved using the Boris method (black plus signs) possess both a drift component and a gyration motion.
The drift velocities, shown in the bottom left panel, are obtained from a linear fit through the particle's $y$-coordinate as a function of time $y(t)$ for GCA, while for the Boris method we remove the gyration by fitting a line through the particle $y$-coordinate minima.
As expected, the accuracy of GCA gradually reduces as the gyroradius increases.
This is quantitatively expressed in the right bottom panel where we plot the relative error of the GCA drift velocity with respect to the (averaged-out) Boris drift, again as a function of the initial velocity and for a constant time step (black plus signs).
The error is $\sim 1\%$ at $R_L/L\approx u_0 = 0.22$ and grows up to $\sim 10\%$ at $R_L/L\approx u_0 = 0.4$.
We also repeated the same test by considering a variable time step (typically much larger), confirming that the errors (circles) remains practically unaltered regardless of the time step size.
Note also that the particle energy cannot change since the right hand side of Eq. (\ref{eq:GCA_final_dupardt}) is trivially zero.

In terms of computational efficiency, the $3^{\rm rd}$-order predictor-corrector method (AM3, Eq. \ref{eq:AM3}) and the explicit method (AB3) are, respectively, $\sim 3.5$ and $\sim 2.2$ more expensive than the Boris method.
Nevertheless, while the Boris scheme requires a time step $\lesssim 0.1$ to accurately sample the Larmor time-scale, GCA integration is only limited by the cell size crossing condition (Eq. \ref{eq:GCA_dt}) yielding a nominal time step of $\sim 100$.

\subsection{Curvature Drift}
\label{sec:curv_drift}

We now turn our attention to the curvature drift by considering a magnetic field configuration in which the only contribution to the Lorentz force comes from magnetic tension, namely
\begin{equation}\label{eq:Bcurv}
  B_R(R)    =  0,\qquad
  B_\phi(R) =  B_0\frac{kR}{\sqrt{1 + k^2R^2}} ,\qquad
  B_z(R)    =  B_0\frac{1} {\sqrt{1 + k^2R^2}}\,, 
\end{equation}
so that  $B^2 = {\rm const}$.
Eq. (\ref{eq:Bcurv}) describes a helical field in cylindrical coordinates $(R,\phi,z)$ with constant pitch $1/k$.
We also include the additional effect of a purely radial electric field,
\begin{equation}
  E_R(R) = \frac{R}{R_0} E_0 \,.
\end{equation}
We employ $N_p = 4$ particles, initially placed on a circle of radius $R_0=100$ with angular position $\phi = j/N_p$ and  four-velocity $u_R=0$, $u_\phi = (j+1)/N_p$, $u_z = 0.25$, where $0\le j < N_p$.
At this distance, the vertical magnetic field is negligible compared to the azimuthal component.
Particles are evolved until $t=2\times 10^3$ by solving the equations of motion in Cartesian coordinates on the computational box defined by $x,y\in[-120,120]$, $z\in[-10,110]$ covered by $128\times128\times64$ grid zones.
We set $B_0=k=e/mc=c=1$ while two values of the electric field are considered, $E_0 = 0$ and $E_0=0.03$.
The time step is held fixed to $\Delta t = 0.1$ (for the Boris method) and $\Delta t = 10$ (for GCA).

\begin{figure*}[!ht]
  \centering
  \includegraphics[width=0.45\textwidth]{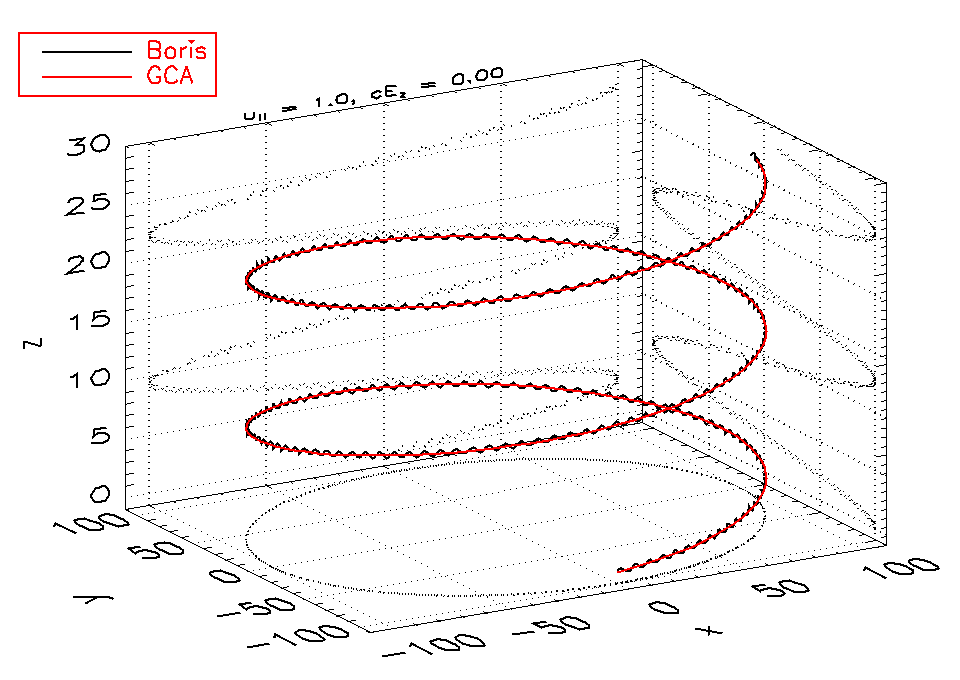}%
  \includegraphics[width=0.45\textwidth]{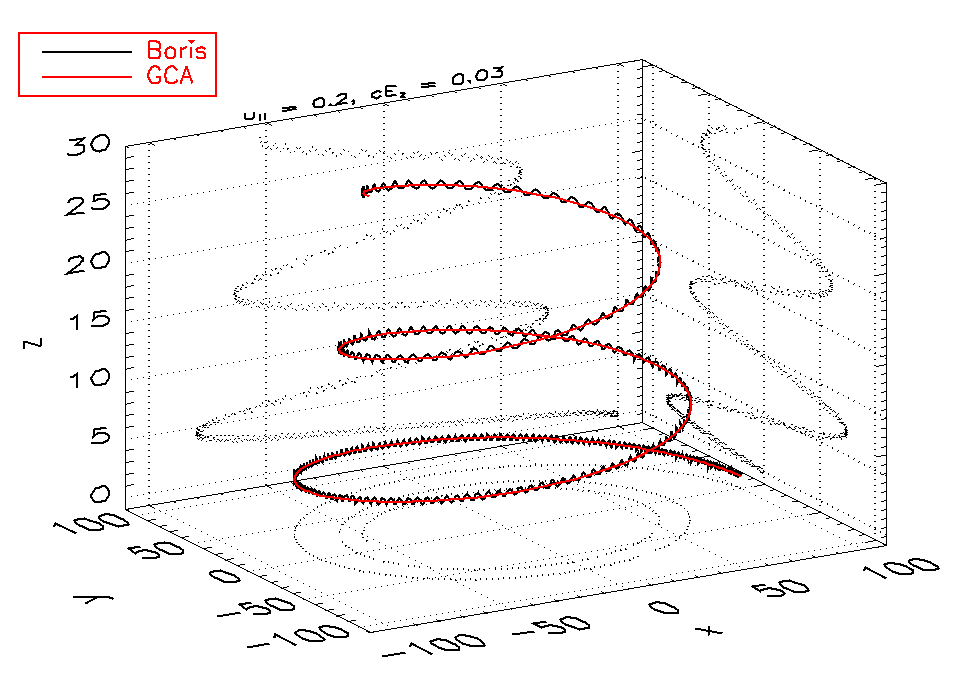}
  \includegraphics[width=0.42\textwidth]{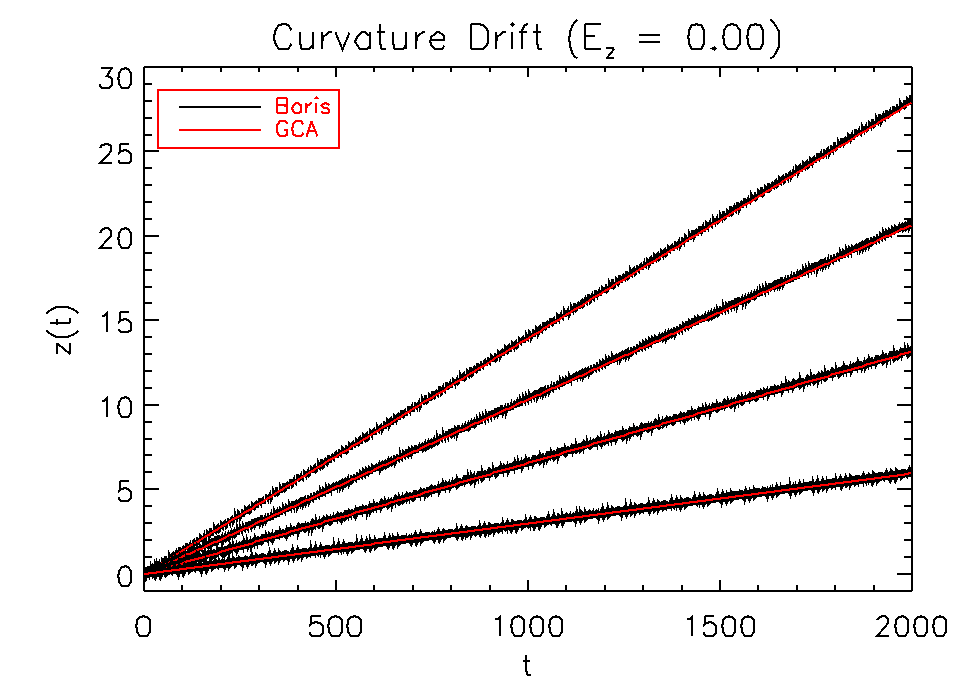}%
  \includegraphics[width=0.42\textwidth]{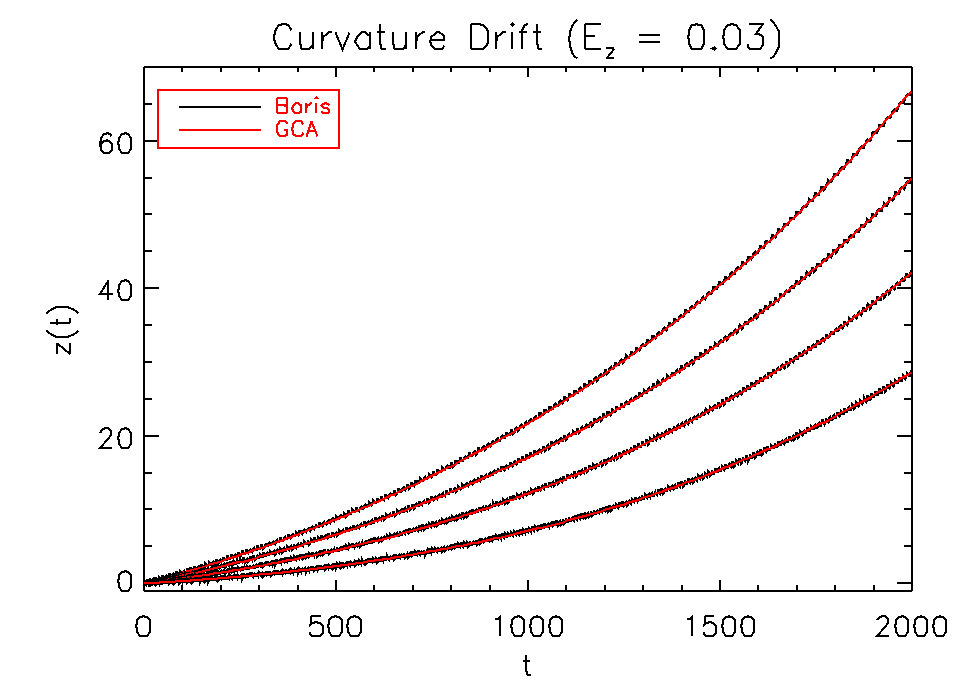}
\caption{\footnotesize
  Particle trajectories (top panels) and vertical position as a function
  of time (bottom panels) for the curvature drift test problem.
  Black and red lines correspond to results obtained with the 
  Boris and GCA scheme, respectively.
  Results on the left (right) have been obtained with $cE_z = 0$ ($cE_z = 0.03$).
  }
  \label{fig:curv_drift}
\end{figure*}
The top panels in Fig. \ref{fig:curv_drift} show the 3D trajectories of two particles with different initial velocities ($u_\parallel = 1$ and $u_\parallel=0.1$) obtained, respectively, when the electric field is zero (left) or $E_0 = 0.03$ (right).
In both cases, the solutions obtained with the GCA (red line) closely overlap with the full orbits recovered by the Boris method (black line).
In the zero electric field case, from Eq. (\ref{eq:GCA_final_dRdt}), one can easily verify that the non-vanishing terms are the parallel velocity $v_\parallel\vec{b}$ and the curvature drift velocity $(\gamma v_\parallel^2/\omega_0) \hvec{b} \times \vec{\kappa}$, where $\vec{\kappa} = (\hvec{b}\cdot\nabla)\hvec{b} \approx -\hvec{e}_R/R_0 $ is the magnetic field line curvature.
The expected vertical position of the guiding center should thus move with approximately constant velocity, $v_{zC} \approx v_\parallel [ b_z - u_\parallel b_\phi/(\omega_0R_0)] $.
This is confirmed in the bottom left panel where we plot the particles vertical position as a function of time obtained with the two methods.
The slope obtained through a linear fit of the $z$-coordinate yields $v_{zC} \approx ( 1.398,\, 1.035,\, 0.659,\, 0.298)\times 10^{-2}$ for GCA and $v_{zC} \approx (1.406,\, 1.040,\, 0.663,\, 0.300)\times 10^{-2}$ for Boris.
As expected, the curvature drift decreases with lower parallel velocities.

The effect of a non-zero radial electric field is twofold. 
On the one hand, it produces an additional perpendicular drift in the radial direction, thus leading to spiral orbits (see top right panel).
On the other, it enhances the curvature drift by increasing the vertical velocity. 
However, while particles gradually shift to lower radii, the velocity does not remain constant as the ratio $B_z/B_\phi \sim 1/(kR)$ becomes larger.

Lastly, we have verified that the particles energy gain is essentially the same for both the Boris scheme and the GCA method, even though the time step of the latter is $100$ larger than the former.

\subsection{X-point}
\label{sec:X_point}

Following \cite{Mori1998}, we investigate the dynamics of charged particles by considering a X-point static configuration given by 
\begin{equation}\label{eq:B_null_field_simple}
    \vec{B} = B_0 \left(\frac{y}{L}, \frac{x}{L}, \frac{B_z}{B_0} \right),
\end{equation}
where $B_0=1$, while $B_z$ and $L$ (the guide field and the system scale, respectively) depend on the specific configuration.
This field shape typically occurs in reconnecting current sheets which are believed to be efficient sources of non-thermal particles in several high-energy astrophysical environments, such as  Active Galactic Nuclei (AGNs; see \cite{Nathanail2020}) and Blazars (\cite{Morris2019}) as well as solar wind (see e.g. \cite{Gosling2012}).
Notice that, in absence of a guide field ($B_z=0$), the GCA equations of motion (\ref{eq:GCA_final_dRdt}) and (\ref{eq:GCA_final_dupardt}) present a singularity at $(x,y)\sim (0,0)$ where the gyroradius (Eq. \ref{eq:RL}) becomes arbitrarily large and the approximation loses its validity.

We assess the accuracy of the GCA by considering two cases corresponding, respectively, to $\vec{E}=\vec{0}$ and $\vec{E}\ne\vec{0}$.

\paragraph{$\vec{E}=\vec{0}$ Case}

\begin{figure*}
    \centering
    \includegraphics[width=0.48\textwidth]{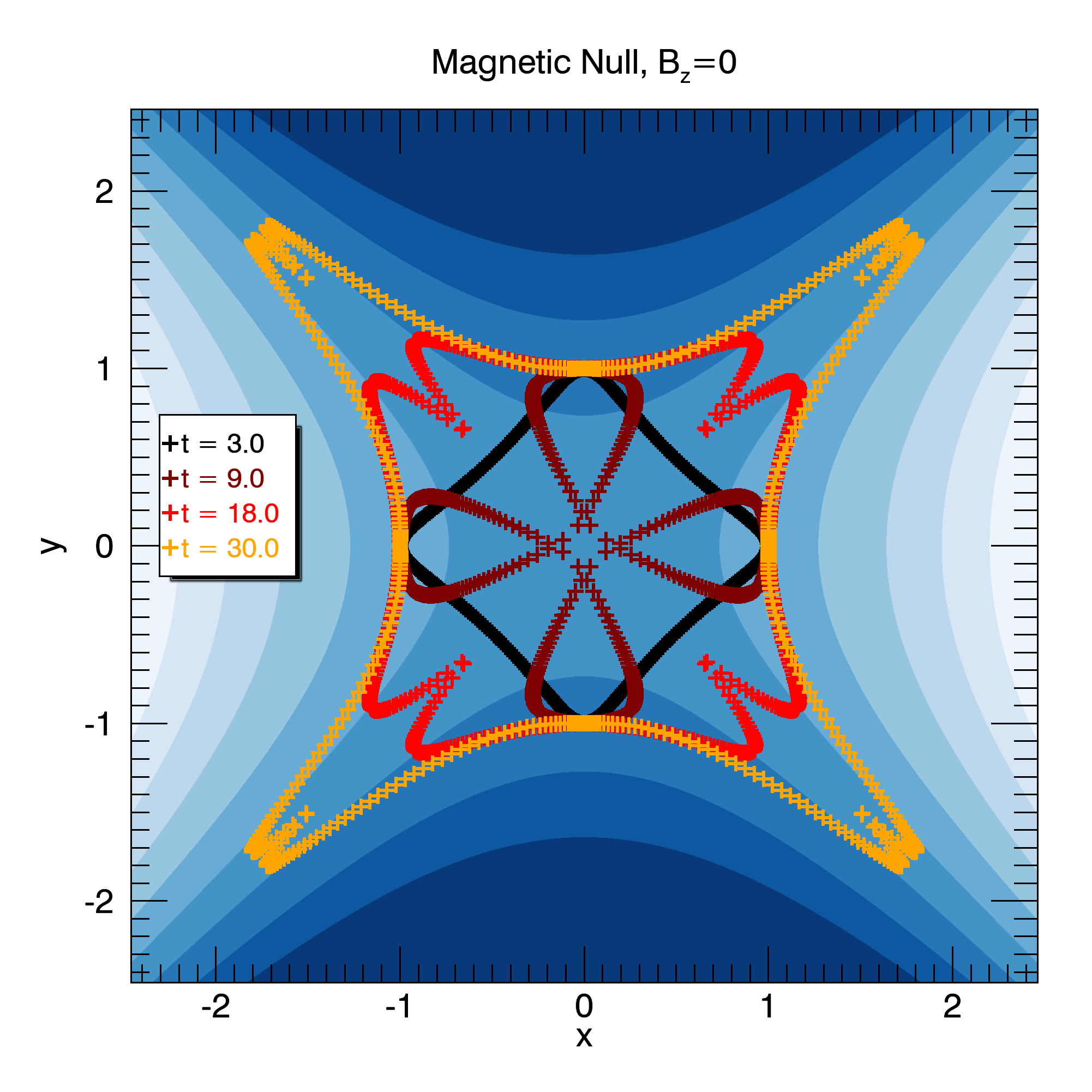}%
    \includegraphics[width=0.48\textwidth]{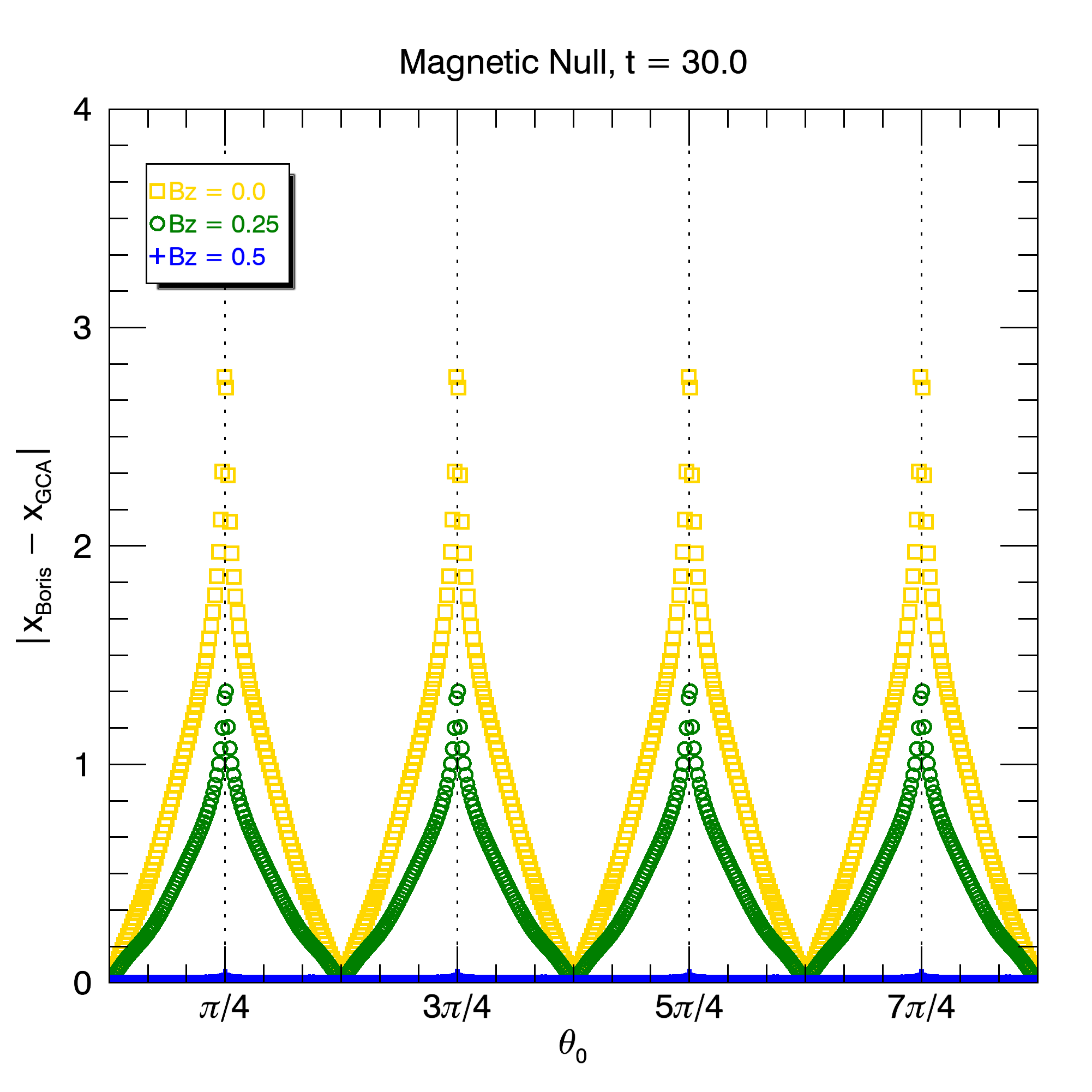}
    \caption{\footnotesize Magnetic null. 
    \textit{Left panel:} particles position colored by time for the magnetic null problem. 
    \textit{Right panel:} absolute errors in the particle position (as a function of the initial angle) for different values guide fields, $B_z = 0$ (yellow), $B_z = 0.25$ (green) and $B_z = 0.5$ (blue).}
    \label{fig:Bnull_xerror}
\end{figure*}

In the first configuration, we set $L=1$ and initialize $500$ equally-spaced particles along the unit circle at $z=0$ with purely radial initial velocity $v_r = -0.1$ pointing towards the origin.
No electric field is present.
As in \cite{Ripperda_etal2018} we consider the cube $[-2.5,2.5]^3$ covered with  $64\times64\times16$  zones and evolve the system until $t_e = 30$ with fixed time step $\Delta t = 5\times10^{-3}$.
Unlike \cite{Ripperda_etal2018}, however, we employ $e/mc=100$ in order to make the Larmor radius smaller compared to $L$ ($R_L\approx 10^{-3})$.
This will make the GCA more appropriate for this test.

Results are shown in Fig. \eqref{fig:Bnull_xerror} for different values of the guide field, $B_z=0, 0.25$ and $0.5$.
Particle trajectories are affected by the relative orientation between the initial velocity and the magnetic field vectors.
Particles initially lying over one of the coordinate axes gyrate and drift perpendicularly to the field lines in the $z$-direction.
Conversely, particles located near the diagonals are very weakly deflected and can approach the domain center because their velocity is initially parallel to the field.
This conclusion holds also for non-zero values of the guide field.
However, regions with $B_z = 0$ (magnetic null) are critical for the GCA, since many terms inside Eq. (\ref{eq:GCA_final_dRdt})-(\ref{eq:GCA_final_dupardt}) grow indefinitely, leading to a loss of accuracy for the GC.
This is confirmed by inspecting the right panel of Fig. \ref{fig:Bnull_xerror}, where we plot the absolute error $|\vec{x}_{\rm Boris} - \vec{x}_{\rm GCA}|$ at the final time as a function of the initial angle. 
A further validation of the null point being the source of precision loss in this test is the considerable reduction of the overall error with the presence of non-zero guide fields (green and blue symbols).

\paragraph{$\vec{E}\ne\vec{0}$ Case}

\begin{figure*}[!ht]
  \centering
  \includegraphics[width=0.49\textwidth]{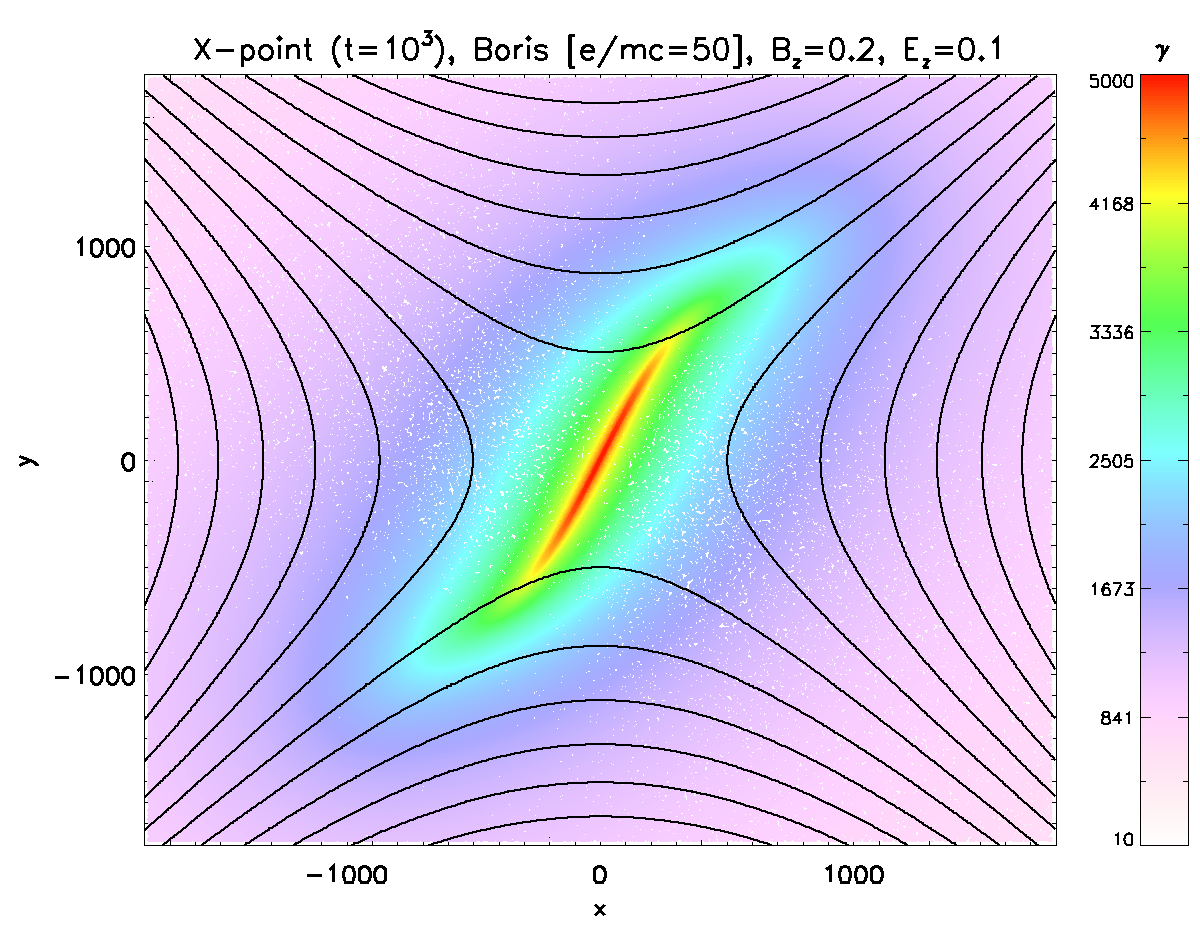}%
  \includegraphics[width=0.49\textwidth]{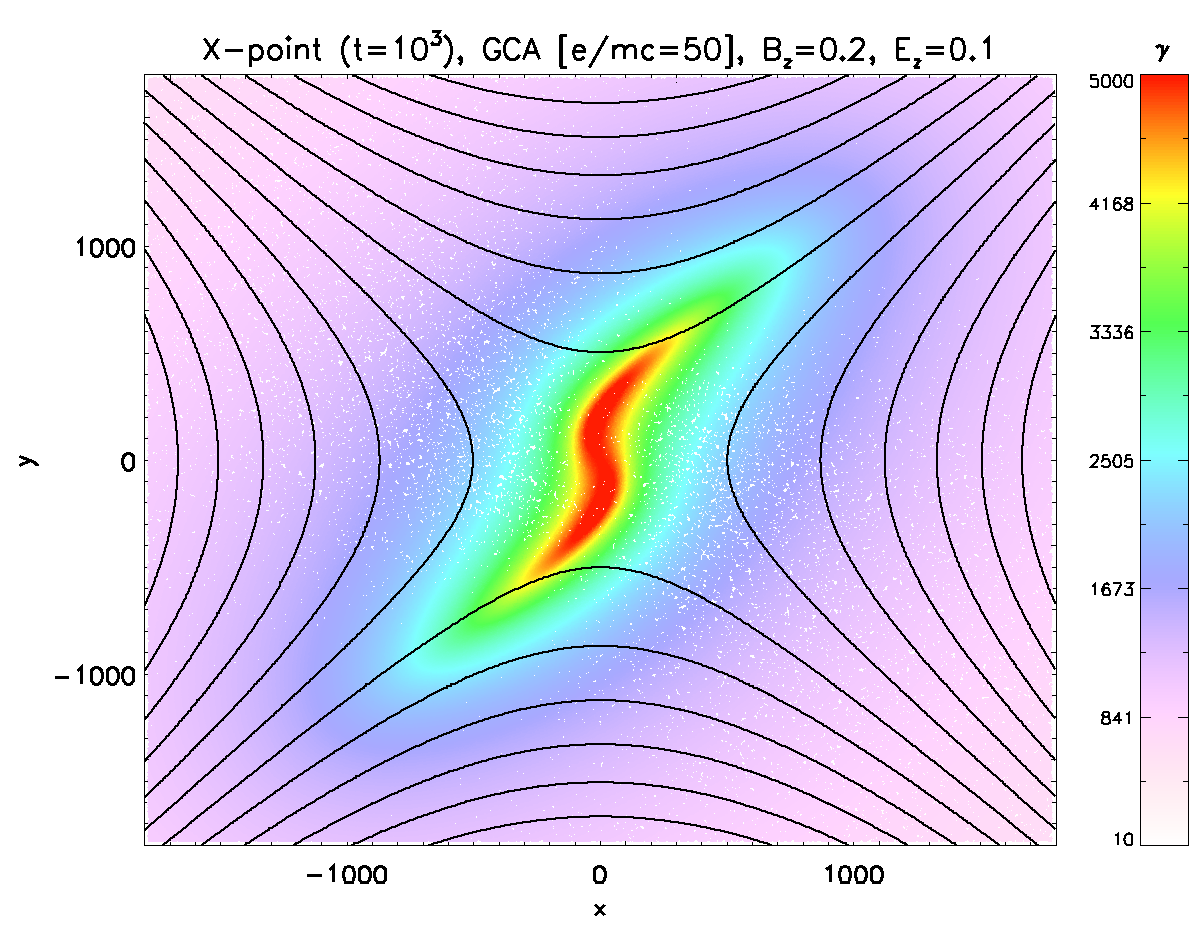}
  \includegraphics[width=0.49\textwidth]{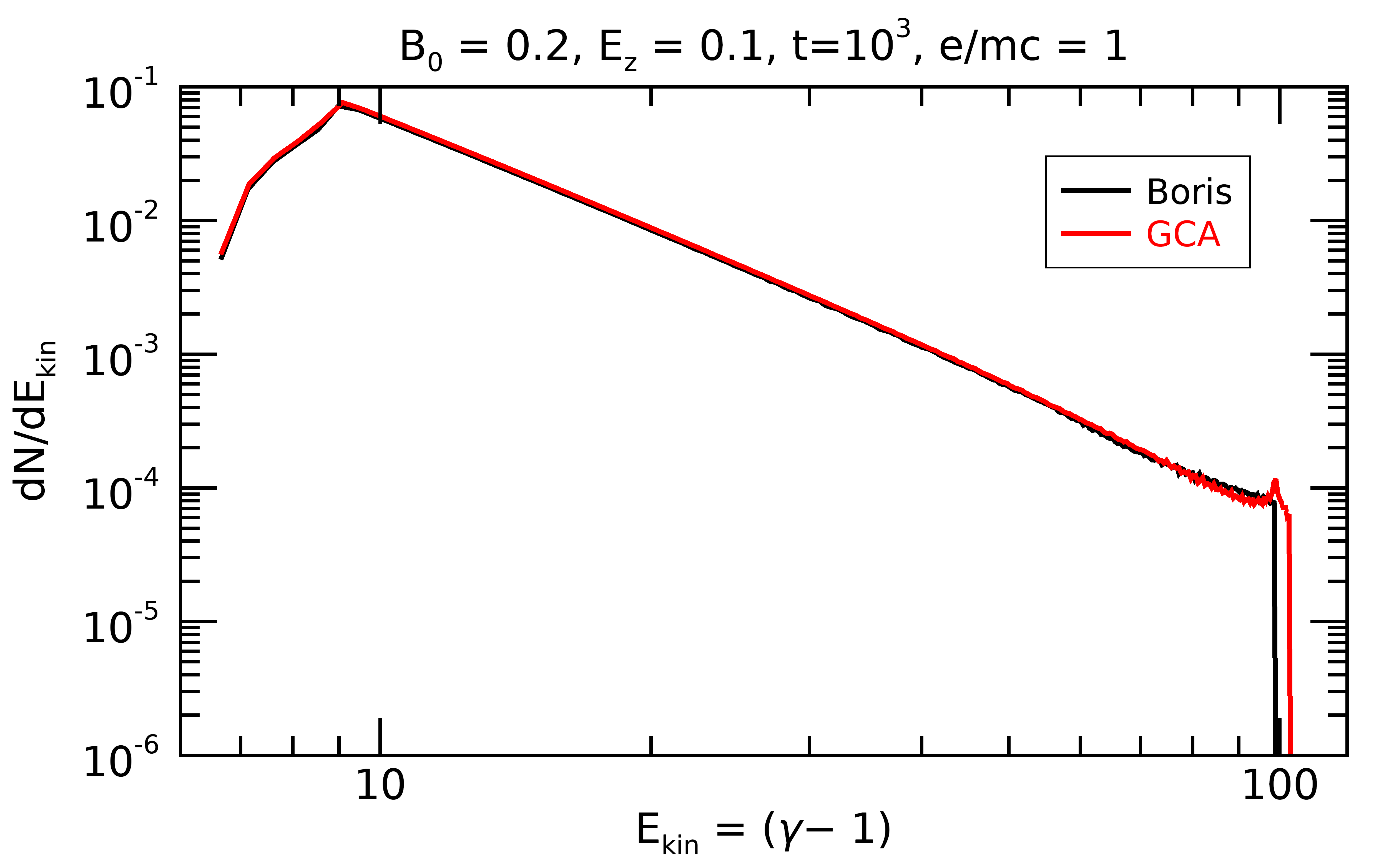}%
  \includegraphics[width=0.49\textwidth]{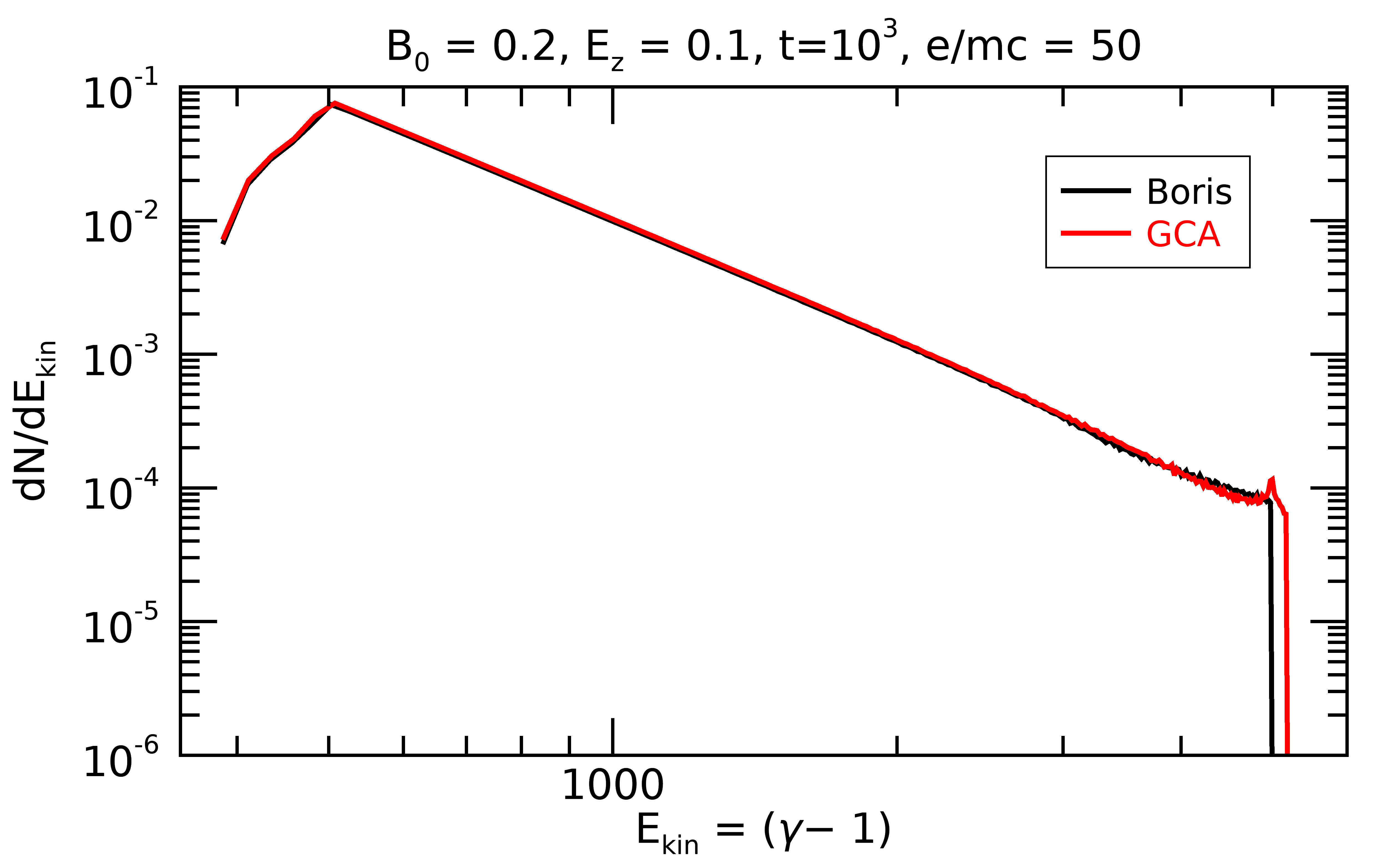}
  \caption{\footnotesize X-point test. 
  \textit{Top panels:} Spatial distributions at $t=10^3$ colored by $\gamma$ Lorentz factor with the Boris (left) and GCA (right) method.
  \textit{Bottom panels:} Comparison of the spectral distribution between Boris and GCA for $e/mc=1$ (left) and $e/mc=50$ (right). 
    }
    \label{fig:tests_Xpoint_emc}
\end{figure*}
Similarly to \S 4.6 of \cite{Mignone_etal2018}, we now introduce a static, uniform electric field $\vec{E} = (0, 0, E_0)$ and employ $1024\times1024$ uniformly spaced zones centered on the origin with domain size $L = 8000$ in both directions.
Slab symmetry is considered along the z-axis. 
We assign $4$ particles per cell using a Maxwellian four-velocity distribution with a thermal velocity corresponding to one-tenth of the speed of light, $v=0.1$.
To avoid loss of validity in the GCA equation, we restrict our attention to configurations with a guide field $B_z\ne0$, in order to ensure that $E/B < 1$ is respected everywhere.
Integration stops at $t=10^3$ using a variable step size.

The particle spatial distribution at $t=10^3$, colored by Lorentz factor,  is shown in the top panels of Fig. \ref{fig:tests_Xpoint_emc}. 
Particles move under the combined effects of perpendicular, gradient and curvature drifts (see \S 4.6 of \cite{Mignone_etal2018}) while favorable acceleration conditions take place in the central region, where $E_z > B_\perp$. 
Energetic particles arrange on an elongated strip approximately lying along the separatrix line $y=x$, which is determined by the sign of the parallel components of the electric and magnetic fields.
With the Boris scheme (top left panel), the shape of the strip agrees with the results of \cite{Mignone_etal2018}  while a slightly more pronounced \quotes{S}-shaped morphology is observed with the GCA (top right panel).
This discrepancy is likely to be attributed to the violation of the nearly crossed field condition (Eq. \ref{eq:crossed_em_condition}), as in the central region of the domain $\vec{E} \sim \vec{E}_\parallel \sim \vec{B}/2$.
The condition implied by Eq. (\ref{eq:crossed_em_condition}), in fact, allows to neglect higher-order terms in the equation of motion for the guiding center (see, for instance, \S 12B in \cite{Vandervoort1960}).
However, if $\vec{E}_\parallel$ becomes comparable to $\vec{B}$, $2^{\rm nd}$-order terms may not be negligible anymore and the GCA results in a loss of accuracy.
Indeed, additional tests (not reported here) confirm that these differences disappear for larger values of the guide field.

The resulting energy spectral distributions, normalized to unity, for $B_z = 0.2$ and $E_0 = 0.1$, are shown in the bottom panels of Fig. \ref{fig:tests_Xpoint_emc} for $e/mc = 1$ (left) and $e/mc = 50$ (right).
The power-laws obtained with the standard Boris scheme and the GCA are comparable, with very small differences, and the maximum energy roughly scales with the charge to mass ratio.
During the initial stages of the simulation, stability of the Boris scheme limits the time step to $\Delta t\sim 7\times 10^{-2}\, mc/e$, while for GCA we have $\Delta t\approx 28$ regardless of $mc/e$.
This offers a tremendous advantage in terms of CPU time even if our GC implementation is roughly $\sim 2.2$ slower to advance a particle when compared to the Boris method.


\begin{figure*}
  \centering
  \includegraphics[width=0.48\textwidth]{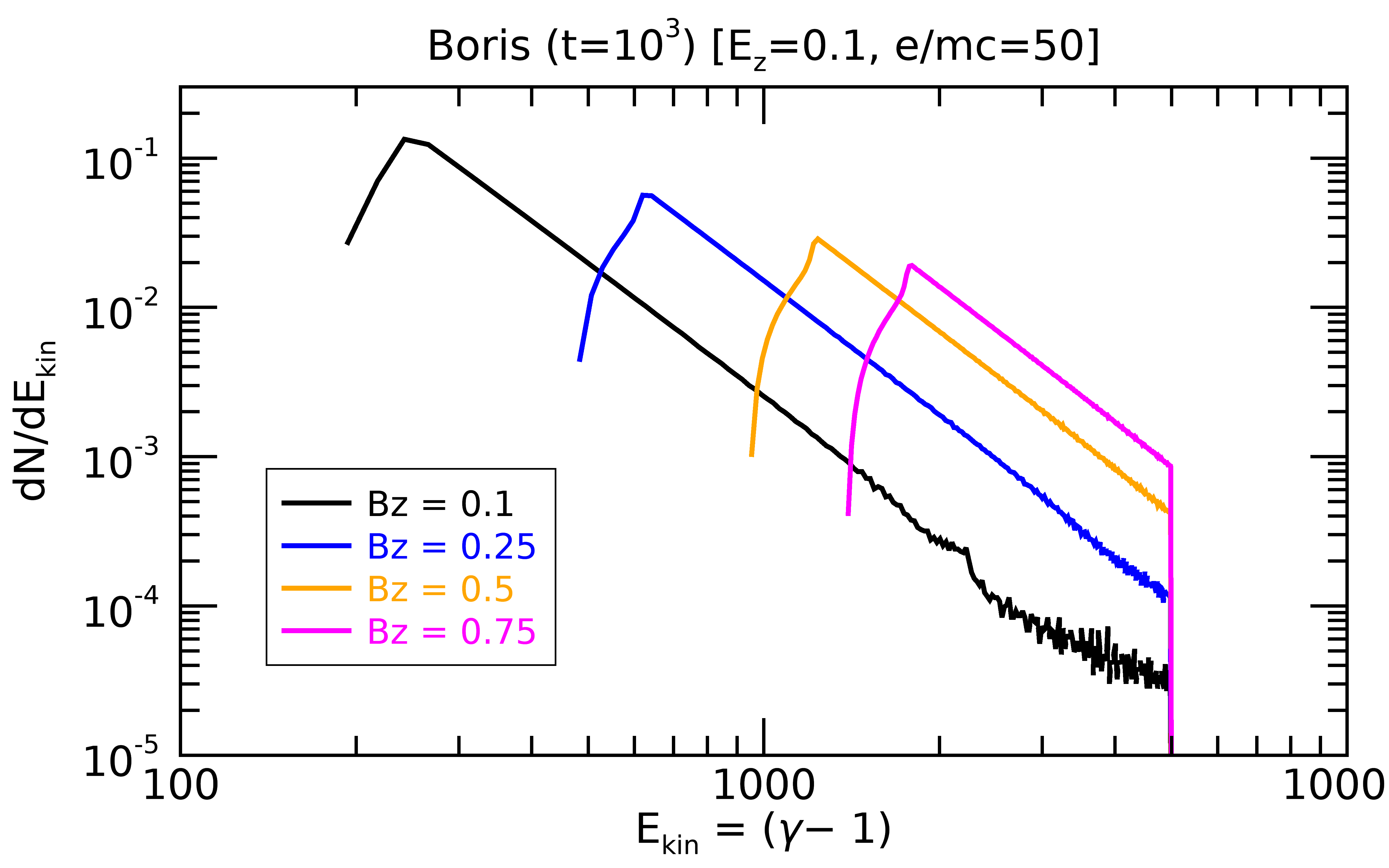}%
  \includegraphics[width=0.48\textwidth]{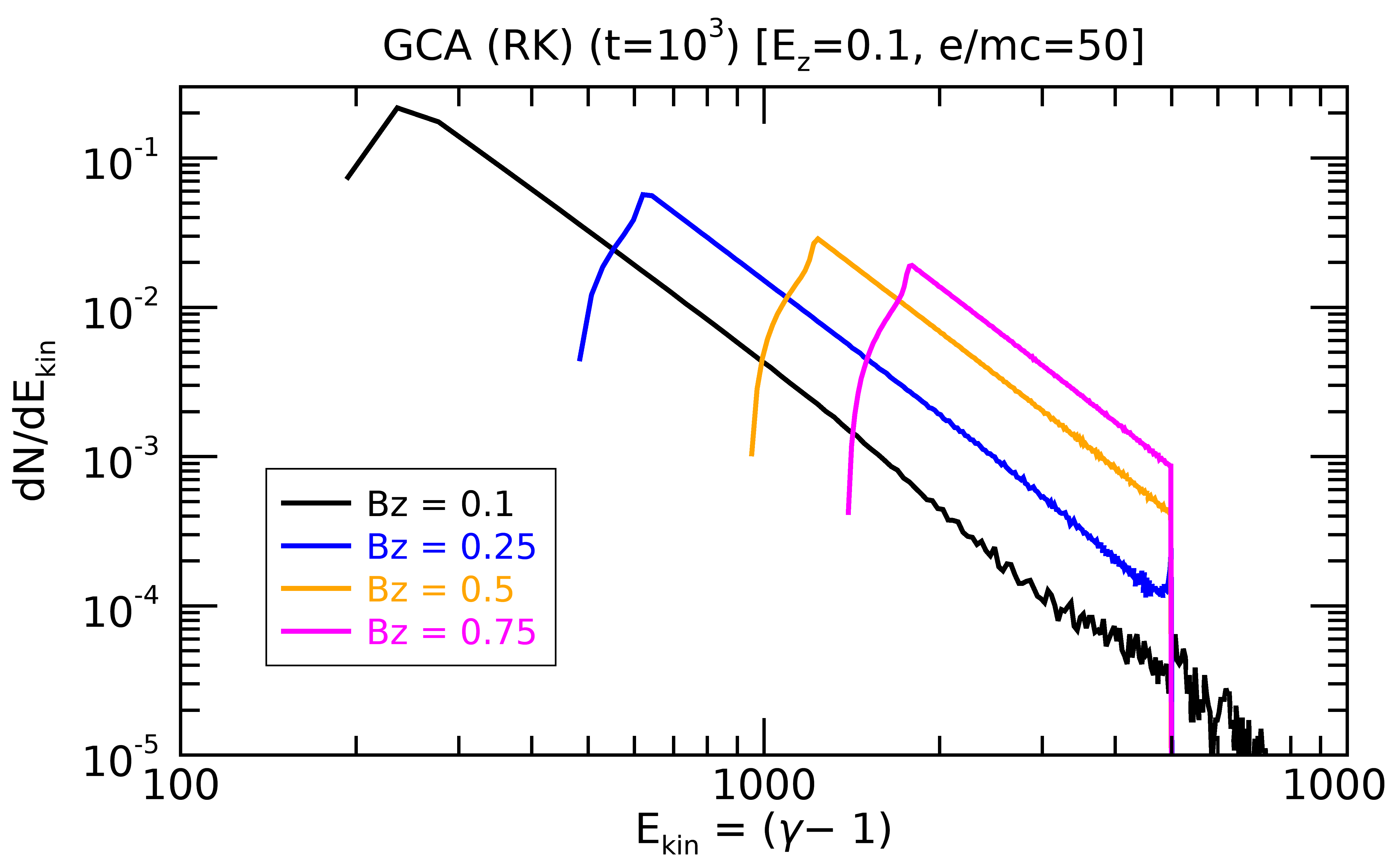}
    \caption{\footnotesize X-point test comparison for the Boris method (left) and GCA (right) with different guide fields, under the same electric field and charge-to-mass ratio. }
    \label{fig:xpoint_Bzs}
\end{figure*}

Last, we consider the influence of different guide fields under the same charge-to-mass ratio $e/m = 50$ and electric field $E_z = 0.1$ in Fig. \ref{fig:xpoint_Bzs}.
The spectra feature a low-energy exponential cutoff, followed by a power-law $\sim \gamma^{-p}$ with index $2 \lesssim p \lesssim 3$ and a sharp cutoff at high energies, in accordance with the results of \cite{Mignone_etal2018}.
As the value of $B_z$ grows, parallel acceleration increases ($\vec{E}\cdot\vec{B}\ne0$ everywhere) and becomes significant for all particles, including those initially away from the null point. 
This causes a systematic shift of the spectral distribution to larger energies while leaving the high-energy cutoff at $\gamma \simeq 500$ unaltered  \citep{Mignone_etal2018}.
As a result, the spectra retain their shape but become narrower for larger values of the guiding field.
The largest acceleration takes place near the X-point.
The most significant difference appears when $B=E=0.1$ (guiding field equal to the electric field): here particles near the origin cross regions with a very small magnetic field (comparable to $\textbf{E}$), their gyration radius increases, and,  as a consequence, the GCA approach becomes less accurate.

\section{Summary}
\label{sec:summary}
%
%

The numerical implementation of the guiding center approximation (GCA) for the PLUTO code for astrophysical plasma dynamics (\cite{Mignone_etal2018}, \cite{Mignone2007}) has been the subject of this work.
The method, originally developed by Northrop \cite{Northrop1963} to describe the relativistic motion of charged particles, is suited for slowly varying fields in which gyration around magnetic field lines is small enough to be neglected but the motion of the gyration center can still be considered as representative of the main particle drifts.
Such conditions are typically met by high energy particles in a variety of highly magnetized plasma, thus of great interest from an astrophysical point of view.
The main strength of the GCA resides in possibility of taking larger computational time steps, since the resolution constraint due to the gyro-period is not needed.

The GCA consists of four ordinary differential equations for the evolution of the guiding center coordinates and the parallel velocity. 
Similarly to particle-in-cell codes, electromagnetic fields and their derivatives are interpolated from the mesh at the particle position.
Due to the large computational overhead, our method is based on a predictor-corrector linear multi-step method with variable step size.
More specifically, we employ the $2^{\rm nd}$-order Adams-Bashforth scheme during the predictor step and the $3^{\rm rd}$-order Adams-Moulton method for the corrector stage.
This combination requires extra storage of the time-derivative (i.e. the right hand side) values at the previous stage and it offers an attractive and more efficient alternative to the traditional Runge-Kutta schemes.

Our implementation results have been validated against and compared to those obtained with the traditional Boris scheme through a selected suite of numerical benchmarks. 
Using the GCA, the particle trajectories in the presence of various drifts (e.g. perpendicular, gradient and curvature drifts) are well reproduced and errors are within the expected approximation, inasmuch the slow-varying condition is respected.
Our predictor-corrector scheme is $\sim 3.5$ times more CPU intensive than the Boris scheme albeit the scheme permits much larger time step, specially for large charge to mass ratios (e.g. electrons) and/or larger magnetizations.
Critical behavior has also been observed - as one would infer from theory - at magnetic null points or in regions where the Larmor radius becomes comparable to the overall field fluctuations.
In these regions, the GCA leads to large errors and its employment becomes unreliable.

As an application example, we have considered particle acceleration in an X-point configuration under the influence of different guide magnetic fields and  nonzero electric fields.
These environments are of great interest in the context of magnetic reconnection and a naturally critical scenario for the GCA due to the abrupt change in the magnetic field over small scales.
A comparative study revealed that GCA consistently reproduces the particles spatial and energy distributions obtained with the Boris method but at a much cheaper numerical cost ($\sim 10^2$), specially during the first evolutionary stages.
Also, in both cases, larger guide fields lead to a more efficient acceleration over the whole domain.

Summarizing, our implementation confirms that the GCA can be an extremely powerful tool to decrease simulation cost, to the extent permitted by the approximation limit.
Its availability in the PLUTO code will hopefully be of great interest to push forward integration times, especially in the field of particle acceleration.
Future extensions of this work will consider hybrid approaches to overcome the critical configurations and inclusion of time-dependent terms.

\section*{Acknowledgments}
The authors wish to acknowledge the University of Torino Scientific Computing Competence Center for the availability of high-performance computing resources and support through the OCCAM cluster.

\appendix
\section{Derivation of GCA equations of motion}
\label{sec:GC_eom_derivation}

In this section we will provide the essentials points in the derivation of the guiding center {\bf equations of motion (EOM)}.
For a complete discussion refer to \cite{Vandervoort1960}.
The equation of motion for a charged particle in an EM field are written in a tensorial form as
\begin{equation}
    \frac{d^2 x_\mu}{d\tau^2} = F_{\mu\nu} \frac{dx_\nu}{d\tau}
        \label{eq:eom_particle} \, ,
\end{equation}
being $x_\mu$ the particle four-coordinate, $F_{\mu\nu}$ the EM tensor and $\tau$ the particle proper time.
The general solution is a linear combination of the two fundamental solutions related to the four EM tensor eigenvalues $q = \pm i \omega$ and $q = \pm \lambda$, where
\begin{align}\label{eq:omega_full}
    &\omega = \frac{e}{mc} \sqrt{ \frac{1}{2} \left(B^2 - E^2\right) + \frac{1}{2} \sqrt{ \left(B^2 - E^2\right)^2 + 4\left(\textbf{E}\cdot\textbf{B}\right)^2}}\,\, , \\
    &\lambda = \frac{e}{mc} \sqrt{ -\frac{1}{2} \left(B^2 - E^2\right) + \frac{1}{2} \sqrt{ \left(B^2 - E^2\right)^2 + 4\left(\textbf{E}\cdot\textbf{B}\right)^2 } } \,\, .
\end{align}
$\omega$ is a generalized relativistic form of the Larmor frequency $\omega_\xi = eB/(mc)$ in the case where $E \neq 0$ (this can be easily verified by solving the limit $E \to 0$). 
The general solution $x_\mu$ of \eqref{eq:eom_particle} is
\begin{equation}
    x_\mu = \xi_\mu \varrho \cos(\omega \tau) - \eta_\mu \varrho \sin(\omega \tau) + \alpha_\mu \nu \cosh(\lambda \tau) + \beta_\mu \nu \sinh(\lambda \tau).
\end{equation}
Note that the first two terms are related to a periodic motion, that is the gyration around magnetic field lines.
Here $\varrho, \nu$ are constants defined by the initial conditions and $\xi_\mu, \eta_\mu, \alpha_\mu$ and $\beta_\mu$ are four-versors normalized in the manner
\begin{equation}
    \xi_\mu^2 = \eta_\mu^2 = 1 \qquad \qquad \textrm{and} \qquad \qquad \alpha_\mu^2 = -\beta_\mu^2 = 1 \, .
\end{equation}
Solutions for $\varrho=0$ and $\nu = 0$ must be valid separately, so by replacing $x_\mu$ for these two particular cases inside Eq. \eqref{eq:eom_particle} we can see that $\xi_\mu, \eta_\mu, \alpha_\mu$ and $\beta_\mu$ form an orthogonal set of the Minkowsky space.
The orthogonality relations between this set of four-vectors in particular states that the periodic motion (gyration) in the ($\xi_\mu,\eta_\mu$)-plane is perpendicular to the acceleration motion in the ($\alpha_\mu,\beta_\mu$)-plane, and will be useful later.
At this point{\BLUE ,} another useful relation is retrieved from the squared velocity invariance
\begin{equation}
    \frac{dx_\mu}{d\tau}\frac{dx_\mu}{d\tau} = - c^2 \,\, , \qquad \textrm{which leads to} \qquad
    \omega^2 \varrho^2 - \lambda^2 \nu^2 = - c^2 \,\, .
    \label{eq:GCA_gammavarrho}
\end{equation}
Since we are only interested in cases in which gyration appears, the only singular eigenvalue case we must consider is the one with $\lambda = 0 $, $\omega \neq 0$, corresponding to $\vec{E} \cdot \vec{B} = 0$ and $|\textbf{E}| \to 0$. 
The associated solution is 
\begin{equation}
    x_\mu = \xi_\mu \varrho \cos(\omega \tau) - \eta_\mu \varrho \sin(\omega \tau) + U_\mu \tau \,\, .
    \label{eq:GCA_eom_particularsolution}
\end{equation}
Here the gyration motion with Larmor frequency $\omega_\xi = \omega$ is clearly shown, along with a uniform motion (drift) with velocity $U_\mu = \left( \gamma c, \vec{U} \right)$ in the plane perpendicular to the $(\vec{E}, \vec{B})$-plane.
For verification simply substitute $x_\mu$ and $U_\mu$ in the left and right-end side of Eq. \eqref{eq:eom_particle}{\BLUE ,}respectively {\BLUE ,} to find that $F_{\mu\nu}U_\nu=0$, and rewrite the indexes of such equation explicitly.
In this particular case, the squared velocity invariance \eqref{eq:GCA_gammavarrho} becomes
\begin{equation}
    \omega^2 \varrho^2 + U_\mu^2 = - c^2 \qquad \textrm{or} \qquad U_\mu^2 = - c^2 - \omega^2 \varrho^2 < - c^2 \,\, .
    \label{eq:GC_fourvel}
\end{equation}
The last equation is of crucial importance in the GCA formalism, because the drift motion $U_\mu$ is the \textit{mean} motion separated from the gyration and corresponds to the GC motion itself, meaning that what we just found is a relation between the energy and momentum of a particle guiding center. 
It is very similar to the well-known formula of the squared four-velocity $U_\mu^{'2} =- c^2$, the only difference being an additional term $\omega^2 \varrho^2$.
This suggests interpreting the motion of the GC as the one of a particle located in the center of gyration, performing no gyration and possessing a squared four velocity as stated in Eq. \eqref{eq:GC_fourvel}, in agreement with part of the particle kinetic energy being stored in the gyration motion through the term $\omega^2 \varrho^2$. 
By separating the spatial and temporal parts of $U_\mu$ and rearranging some terms inside Eq. \eqref{eq:GC_fourvel}{\BLUE ,} we can get a relation for the GC Lorentz factor
\begin{equation}
    \gamma_{GC} = \sqrt{1 + \frac{|\textbf{U}|^2}{c^2} + \frac{\omega^2 \varrho^2}{c^2}} = \sqrt{1 + \frac{\gamma^2 |\textbf{v}|^2}{c^2} + \frac{\omega^2 \varrho^2}{c^2}} \, .
\end{equation}
Although we only considered a limit case, the same equation can be proven to be generally valid in the GCA formalism.

We now want to isolate the gyration motion in a general case. 
This is done by separating the motion as follows:
\begin{equation}
    x_\mu = \xi_\mu x + \eta_\mu y + x'_\mu + X_\mu \, ,
\end{equation}
where $x_\mu$ is the space-time position of the particle and $X_\mu$ is the position of the guiding center.
We associate two variables $x$ and $y$ to describe the periodic motion in the $(\xi,\eta)$-plane, and the term $x_\mu'$ will account for the periodic motions which do not lie in the same plane.
To more easily describe the gyration, we also choose a new pair of variables
\begin{equation}
    \zeta = \frac{1}{\sqrt{2}} (x + iy) \qquad \qquad \textrm{and} \qquad \qquad \zeta^* = \frac{1}{\sqrt{2}} (x - iy) \, ,
    \label{eq:eta_eta*}
\end{equation}
so that $\xi_\mu x + \eta_\mu y = \delta_\mu \zeta + \sigma_\mu \zeta^*$. 
The new variables will allow us to choose $\sigma_\mu$ and $\delta_\mu$ in a more convenient way and let some factor $e^{i\phi}$ account for the initial conditions of the problem.
We rewrite the particle coordinate using Eq. \eqref{eq:eta_eta*} 
\begin{equation}
    x_\mu = \delta_\mu \zeta + \sigma_\mu \zeta^* + x_\mu' + X_\mu{\BLUE .}
    \label{eq:GCA_equation_newvariables}
\end{equation}
The next step is substituting this expression into \eqref{eq:eom_particle} and expand the second derivative, which leads to a really long expression containing eight unkowns: $\zeta$ , $\zeta^*$, two components of $x_\mu$ (the condition with the ($\sigma,\delta$)-plane sets the other two), and four components of $X_\mu$.

The conditions for the GCA expressed in Eq. \eqref{eq:GCA_conditions} implicitly contain a parameter of smallness $\epsilon = |u/\omega L| \ll 1$, where $u$ is the typical particle four-velocity and $L$ the scale length upon which the change in the EM fields $F_{\mu\nu}$ is comparable to $F_{\mu\nu}$ (slowly-varying fields approximation). 
This allows to approximate $F_{\mu\nu}(x_\mu)$ (particle coordinate) by using a Taylor expansion about $X_\mu$ (the GC coordinate)
\begin{align}
    \begin{split}
        F_{\mu\nu}(x_\mu) &= F_{\mu\nu}(X_\mu) + \frac{\partial F_{\mu\nu}}{\partial x_\xi} \left(\delta_\xi \zeta + \sigma_\xi \zeta^*\right) \\
        &+ \frac{1}{2} \frac{\partial^2 F_{\mu\nu}}{\partial x_\xi \partial x_\pi} \bigg[ \delta_\xi \delta_\pi \zeta^2 + \sigma_\xi \sigma_\pi \zeta^{*2} \\
        &+ \left( \sigma_\xi \delta_\pi + \sigma_\pi \delta_\xi \right) \left|\zeta^2\right| \bigg]+ \frac{\partial F_{\mu\nu}}{\partial x_\xi} x'_\xi + \cdots
    \end{split}
\end{align}
We should now substitute $F_{\mu\nu}(x_\mu)$ inside Eq. \eqref{eq:eom_particle} using the new coordinates in order to get a quite long equation containing all quantities evaluated at the GC position $X_\mu$ up to $2\nd-$order, and group all $2\nd-$order terms on the right-end side $G_\mu$
\begin{equation}
    \begin{aligned}
            &\delta_\mu \frac{d^2 \zeta}{d \tau^2} + \sigma_\mu\frac{d^2 \zeta^*}{d \tau^2} + \left( i \omega \delta_\mu + 2\frac{d\delta_\mu}{d\tau}\right) \frac{d\zeta}{d\tau}\\
            &+ \left( - i \omega \sigma_\mu + 2\frac{d\sigma_\mu}{d\tau}\right) \frac{d\zeta^*}{d\tau} + \frac{d^2 \delta_\mu}{d\tau^2}\zeta + \frac{d^2\sigma_\mu}{d\tau^2} \zeta^*\\
            &- \frac{\partial F_{\mu\nu}}{\partial x_\xi} \left( \delta_\xi \zeta + \sigma_\xi \zeta^* \right) \frac{d X_\nu}{d\tau} - F_{\mu\nu} \left( \frac{d\delta_\nu}{d\tau} \zeta + \frac{d\sigma_\nu}{d\tau} \zeta^* \right) \\
            &- \frac{\partial F_{\mu\nu}}{\partial x_\xi} \left( \delta_\xi \zeta + \sigma_\xi \zeta^* \right) \left( \delta_\nu \frac{d\zeta}{d\tau} + \sigma_\nu \frac{d\zeta^*}{d\tau} \right)\\
            &+ \left( \frac{d^2 X_\nu}{d\tau^2} - F_{\mu\nu} \frac{dX_\mu}{d\tau} \right) + \left( \frac{d^2 x'_\mu}{d\tau^2} - F_{\mu\nu} \frac{dx'_\nu}{d\tau} \right) = G_\mu \, ,
    \end{aligned}
    \label{eq:GCA_complete_firstorder_G_i}
\end{equation}
An expression for $\zeta$ is needed to further proceed, but since the derivation is quite long we will not explicitly derive it here.
The basic idea is to use the orthogonality equations for the new four-vectors $\sigma_\mu$ and $\delta_\mu$ and the antisymmetry property of the EM tensor inside Maxwell's equations written in tensorial form, and then assume that a solution for $X_\mu$ has already been found.
Considered that $x_\mu$ is a $1\st-$order term in the GCA, by ignoring $2\nd-$order quantities and writing terms independent from $\zeta$ as a numerical value $P$, then Eq. \eqref{eq:GCA_complete_firstorder_G_i} becomes
\begin{equation}
    \frac{d^2\zeta}{d\tau^2} + i\Omega\frac{d\zeta}{d\tau} + \frac{i}{2} \frac{d\Omega}{d\tau}\zeta + \Omega a \zeta + P = 0
    \label{eq:GCA_zeta_eom} \, ,
\end{equation}
where $\Omega a \zeta$ contains all the linear terms in $\zeta$.
Consequently, the corresponding solution $\zeta$, by analogy to the {\bf Wentzel–Kramers–Brillouin (WKB)} approximation in quantum mechanics, is
\begin{equation}
    \zeta = \varrho_0\sqrt{\frac{\omega_0}{\Omega}} e^{-i \left( \Phi + \int_{\tau_0}^\tau\sigma_1 d\tau \right)} \,\, ,
    \label{eq:GCA_zeta_firstorder}
\end{equation}
to the $0^\mathrm{th}-$order, being
\begin{equation}
    \Omega = \omega - 2i\sigma_\mu\frac{d\delta_\mu}{d\tau}\,\, , \,\, \Phi = \int_{\tau_0}^\tau \Omega d\tau - \phi\,\, \textrm{and} \,\, \sigma = a - \frac{a^2}{\Omega} \,\, .
\end{equation}
The subscripts $0, 1$ indicate the associated order of approximation for quantities defined above.

Since now we know the expression for $\zeta$, we can write the $1\st-$order EOM for $X_\mu$.
For this purpose, we require that in the left-end side of Eq. \eqref{eq:GCA_complete_firstorder_G_i} the terms containing the $1\st-$ and $2\nd-$order derivatives of $X_\mu$ are balanced by the remaining nonoscillatory terms, that is
\begin{equation}
    \frac{d^2X_\mu}{d\tau^2} - F_{\mu\nu}\frac{dX_\nu}{d\tau} = \frac{\partial F_{\mu\nu}}{\partial x_\xi} \left( \delta_\xi \sigma_\nu \zeta \frac{d \zeta^*}{d \tau} + \delta_\nu \sigma_\xi \zeta^* \frac{d \zeta}{d \tau} \right) \,\, .
\end{equation}
Having found the solution \eqref{eq:GCA_zeta_firstorder} and making use once again of the orthogonality relations for $\delta_\mu$ and $\sigma_\mu$ and the tensorial Maxwell's equations{\BLUE ,} we can write the equation of motion for the GC
\begin{equation}
    \frac{d^2X_\mu}{d\tau^2} - F_{\mu\nu}\frac{dX_\nu}{d\tau} + \varrho_0^2 \omega_0 \frac{\partial \omega}{\partial x_\mu} = 0 \,\, .
\end{equation}
%

\section{First order solution for the GCA equations of motion}
\label{sec:GCA_eom_solution}
%
%
%

An explicit solution for Eq. \eqref{eq:GCA_GC_eom} can be found under the approximation \eqref{eq:crossed_em_condition}{\BLUE ,} which introduces a new parameter of smallness $\lambda/\omega \ll 1$. 
We will treat the new approximation analogously to the one introduced in the slowly-varying field case, since we are interested in the first order in both cases and the distinction between them would be a merely formal procedure.
As stated in \S \ref{sec:equations}, we express the EM tensor under this approximation as the sum of a $0^{th}-$order tensor, constructed from $\textbf{B}$ and $\textbf{E}_\perp$, and a $1\st-$order tensor, constructed from $\textbf{E}_\parallel$.
In the same way, the GC four-velocity $U_\mu = dX_\mu/d\tau$ is decomposed as $U_\mu = U_\mu\zero + U_\mu\one$ (we remind that the superscript $i$ indicates the $i^{th}$-order of approximation), so that Eq. \eqref{eq:GCA_GC_eom} leads to a system of two equations
\begin{align}
    &F_{\mu\nu}\zero U_\nu\zero = 0{\BLUE ,}
    \label{eq:GC_eom_zeroorder_Fik}\\
    &\frac{dU_\mu\zero}{d\tau} = F_{\mu\nu}\zero U_\nu\one + F_{\mu\nu}\one U_\nu\zero  - \rho_0^2 \omega_0 \frac{\partial \omega}{\partial x_\mu} \,\, .
    \label{eq:GC_eom_firstorder_Fik} 
\end{align}
By using the orthogonality relations between $\xi, \eta, \alpha$ and $\beta$ and this last equations{\BLUE ,} it can be proved that $U_\mu\one$ is composed of a $1\st-$order correction $U_\mu^L$ in the direction of $U\zero$, and a term $U_\mu^P$ perpendicular to the other two. 
Knowledge about $U_\mu^L$ is not useful for most applications of the GCA, so to find an expression for the guiding center four-velocity we will only need to determine $U_\mu = U_\mu\zero + U_\mu^P$.
We rewrite $U_\mu = \left( \gamma c, \textbf{U} \right)= \left( \gamma c, \gamma \textbf{v} \right)$, so that the separated space and time components of Eq.\eqref{eq:GC_eom_zeroorder_Fik} are
\begin{align}
    &\textbf{U}\zero \times \textbf{B} + \textbf{E}_\perp \gamma c = 0 \\
    &\textbf{E}_\perp \cdot \textbf{U}\zero = 0 \, .
\end{align}
Here we can clearly see that the $0^\textrm{th}-$order term has to be 
\begin{equation}
    U_\mu\zero = \left(\gamma c, U_\parallel\zero \textbf{b} + \gamma \textbf{v}_E\right) = \left(\gamma c, \gamma v_\parallel \textbf{b} + \gamma \textbf{v}_E\right){\BLUE ,}
    \label{eq:GCA_U_izero}
\end{equation}
being $\textbf{b}$ the unit vector in the direction of $\textbf{B}$ and $\textbf{v}_E = c \textbf{E}_\perp \times \textbf{B}/B^2 = c\textbf{E} \times \textbf{b}/B$.
This means that to the lowest order the GC possesses a parallel velocity along magnetic field lines, as well as the known $\textbf{E}\times\textbf{B}$ drift in the direction perpendicular to $\textbf{b}$.
Likewise, Eq. \eqref{eq:GC_eom_firstorder_Fik} becomes
\begin{align}
    &\frac{d\textbf{U}\zero}{d\tau} = \frac{e}{mc}\left( \textbf{U}\one\times\textbf{B} + U_0\one\textbf{E}_\perp + \gamma c \textbf{E}_\parallel \right) - \mu_0 \nabla \omega{\BLUE ,}
    \label{eq:GC_dU0dtau}
    \\
    &\frac{d \gamma c}{d\tau} = \frac{e}{mc} \left( \textbf{E}_\perp \cdot \textbf{U}\one + \textbf{E}_\parallel \cdot \textbf{U}\zero \right) + \frac{\mu_0}{c}\frac{\partial \omega}{\partial t} \, .
    \label{eq:GC_dgammadtau} 
\end{align}
Eq. \eqref{eq:GCA_final_dupardt} can be obtained by multiplying the spatial part by $\textbf{b}$ (and remembering the form we found for $\textbf{U}\zero$)
\begin{equation}
    \frac{d (\gamma v_\parallel)}{d\tau} = - \gamma \textbf{b}\cdot \frac{d\textbf{v}_E}{d\tau} + \frac{e}{mc} \gamma c \textbf{b}\cdot \textbf{E}_\parallel - \mu_0 \textbf{b}\cdot \nabla \omega \,\, ,
    \label{eq:GC_upar_eom}
\end{equation}
where some terms were simplified by virtue of the fact that $\textbf{b}$ is orthogonal to both $d\textbf{b}/d\tau$ and $\textbf{v}_E$.
Eq. \eqref{eq:GC_upar_eom} can be used to determine $\gamma v_\parallel$ once we have an expression for $\textbf{U}^P$. 
Since $U_\mu\one$ is perpendicular to the magnetic field, we may take the vector product of Eq. (\ref{eq:GC_dU0dtau}) with $\textbf{b}$.
This suppresses the components in the direction of the field and results in an equation that can be used to determine $U_\mu^P$
\begin{equation}
    \begin{aligned}
        \gamma v_\parallel& \textbf{b}\times \frac{d\textbf{b}}{d\tau} + \textbf{b}\times \textbf{v}_E \frac{d(\gamma)}{d\tau} + \gamma \textbf{b} \times \frac{d\textbf{v}_E}{d\tau}\\ 
        &= \frac{eB}{mc^2} \left\{ \textbf{U}\one - \textbf{b}\left(\textbf{b}\cdot \textbf{U}\one\right) \right\} - \frac{eB}{mc} U_0\one \textbf{v}_E - \mu_0\textbf{b}\times \nabla \omega \, .
        \label{eq:GCA_234}
    \end{aligned}
\end{equation}
We first note that Eq. \eqref{eq:GCA_234} does not have components directed along the magnetic field, since the inhomogeneous terms have components only in the plane perpendicular to $\textbf{b}$.
We then make use of Eq. \eqref{eq:GC_dU0dtau} and \eqref{eq:GCA_U_izero} to rewrite
\begin{equation}
    \begin{aligned}
        \textbf{b}\times\textbf{v}_E & \frac{d(\gamma)}{d\tau} = \frac{E_\perp}{B} \boldsymbol\varepsilon \frac{d(\gamma c)}{d\tau} \\
        &= \frac{eB}{mc} \frac{E_\perp^2}{B^2} \boldsymbol\varepsilon ( \boldsymbol\varepsilon \cdot \textbf{U}^P) + \frac{eB}{mc} \frac{\textbf{E}_\perp}{B^2} \gamma v_\parallel E_\parallel + \frac{\textbf{E}_\perp}{B} \frac{\mu_0}{c}\frac{\partial \omega}{\partial t}{\BLUE ,}
    \label{eq:GCA_238}
    \end{aligned}
\end{equation}
where $\boldsymbol\varepsilon$ is a unit vector directed along $\textbf{E}_\perp$ and we replaced $\boldsymbol\varepsilon \cdot \textbf{U}\one = \boldsymbol\varepsilon \cdot \textbf{U}^P$.
We also make use of the fact that $U_\mu^P$ is orthogonal to $U_\mu\zero$ and its spatial part $\textbf{U}^P$ is perpendicular to $\textbf{b}$, so that the product $U_\mu\zero U_\mu^P = 0$ can be rewritten as
\begin{equation}
    \frac{eB}{mc} \boldsymbol\alpha U_0^P = \frac{eB}{mc^3} \textbf{v}_E(\textbf{v}_E\cdot\textbf{U}^P) = \frac{eB}{mc} \frac{E_\perp^2}{B^2} \boldsymbol\varepsilon \times \textbf{b} \left\{ \left(\boldsymbol\varepsilon \times \textbf{b}\right) \cdot \textbf{U}^P \right\} \,\,.
    \label{eq:GCA_240}
\end{equation}
Now we can substitute Eq. \eqref{eq:GCA_238} and \eqref{eq:GCA_240} inside \eqref{eq:GCA_234}, rearrange some terms and find
\begin{equation}
    \begin{aligned}
        \textbf{U}^P & = \frac{mc\gamma v_\parallel}{e(B^2 - E_\perp^2)}\textbf{B}\times\frac{d\textbf{b}}{d\tau} + \frac{mc\gamma }{e(B^2 - E_\perp^2)}\textbf{B}\times\frac{d\textbf{v}_E}{d\tau} \\
        & + \frac{\gamma v_\parallel E_\parallel}{(B^2 - E_\perp^2)}\textbf{E}_\perp + \frac{mc \mu_0 }{e(B^2 - E_\perp^2)} \left( \textbf{B}\times\nabla\omega + \frac{\textbf{E}_\perp}{c}\frac{\partial \omega}{\partial t} \right) \,\, .
    \end{aligned}
\end{equation}
Finally, because the perpendicular electric field satisfies the relation $\textbf{E}_\perp = \textbf{b}\times (\textbf{E}\times \textbf{b}) = \textbf{B} \times \textbf{v}_E$, we obtain
\begin{equation}
    \begin{aligned}
        \textbf{U}^P & = \frac{\textbf{b}}{B\left(1-\frac{E_{\perp}^2}{B^2}\right)} \times \frac{mc}{e} \Bigg\{ \gamma \left( v_\parallel\frac{d\textbf{b}}{d\tau}  + \frac{d\textbf{v}_E}{d\tau} \right) \\
        &+ \frac{\mu_0}{c^2} \textbf{v}_E \frac{\partial\omega}{\partial t} + \frac{\mu_0}{c} \nabla \omega + \frac{e \gamma v_\parallel E_\parallel}{mc^2}\textbf{v}_E  \Bigg\} \,\, .
    \end{aligned}
\end{equation}
This last expression, together with \eqref{eq:GCA_U_izero}, provides the $1\st$-order velocity of the guiding center $U_\mu$, from which we get Eq. \eqref{eq:GCA_final_dRdt}.




\bibliographystyle{mnras}
\bibliography{bibliography}







\end{document}